\documentclass[%
preprint,
 amsmath,amssymb,
 aps,
]{revtex4-1}

\usepackage[dvipdfmx]{graphicx}% Include figure files
\usepackage{dcolumn}% Align table columns on decimal point
\usepackage{bm}% bold math
\usepackage{amsmath}
\usepackage{amssymb}
\usepackage{mathrsfs}
\usepackage{natbib}
\usepackage{mathtools}

\newcommand{\bepsilon}{\scalebox{0.7}{$\langle$}\epsilon\scalebox{0.7}{$\rangle$}}
\newcommand{\bchi}{\scalebox{0.8}{$\langle$}\chi\scalebox{0.8}{$\rangle$}}
\newcommand{\mathsfbi}[1]{\boldsymbol{\mathsf{#1}}}

\makeatletter
 \renewcommand{\theequation}{\arabic{section}.\arabic{equation}}
 \@addtoreset{equation}{section}
\makeatother

\begin{document}

%\preprint{APS/123-QED}

\title{Hessian-based Lagrangian closure theory\\ for passive scalar turbulence}% Force line breaks with \\
%\thanks{A footnote to the article title}%
\author{Taketo Ariki$^1$}
\email{taketo.ariki.c5@tohoku.ac.jp}
% \altaffiliation[Also at ]{EcoTopia Science Institute, Nagoya University\\
% Furocho, Chikusa-ku, Nagoya, Japan}%Lines break automatically or can be forced with \\
\author{Kyo Yoshida$^2$}
\email{yoshida.kyo.fu@u.tsukuba.ac.jp}
\affiliation{%
$^1$Department of Aerospace Engineering, Tohoku University, 6-6-01 Aramaki-Aza-Aoba, Aobaku, Sendai, Japan\\%\vspace{5mm}
$^2$Division of Physics, Faculty of Pure and Applied Sciences,
University of Tsukuba, 1-1-1 Tennoudai, Tsukuba 305-8571, Japan}%
\date{\today}% It is always \today, today,
             %  but any date may be explicitly specified

\begin{abstract}
Self-consistent closure theory for passive-scalar turbulence has been developed on the basis of the Hessian of the scalar field. As a primitive indicator of spatial structure of the scalar, we employ the Hessian into the core of the theory to properly characterize the time scale intrinsic to the scalar field itself. The resultant closure model is now endowed with several realistic features, i.e. the scale-locality of the interscale interaction, the detailed conservation, and the memory-fading effect. Applying the current theory to the inertial-convective range eventually leads to self-consistent derivation of the Obukhov-Corrsin spectrum with its universal constant consistent with numerical and experimental data.
\begin{description}
\item[PACS numbers] 
%May be entered using the \verb+\pacs{#1}+ command.
%\item[Structure]
%You may use the \texttt{description} environment to structure your abstract;
%use the optional argument of the \verb+\item+ command to give the category of each item. 
\end{description}
\end{abstract}

\pacs{Valid PACS appear here}% PACS, the Physics and Astronomy
                             % Classification Scheme.
%\keywords{Suggested keywords}%Use showkeys class option if keyword
                              %display desired
\maketitle

\section{Introduction}
There are in nature a number of phenomena realizing highly disordered flows of gases and liquids, where turbulence plays indispensable roles in transporting varieties of physical properties; e.g., energy, mass, chemicals, charge, etc. While these transported properties themselves can actively change the flow properties in general, many fundamental aspects can be studied from their passive convection. Thence, passively-convected scalar has been studied as the simplest model of transportation phenomena caused by turbulence, providing a prototypical understanding of turbulence-mixing effect \cite{Warhaft00}. In case of sufficiently high Reynolds and P\'{e}clet numbers, one may observe a well-developed inertial-convective range where scalar statistics represent universal scaling laws irrespective of the large and small-scale mechanisms of both velocity and scalar fields, which is achieved by the classical Kolmogorov-Obukhov-Corrsin theory using a dimensional analysis \cite{K41a,Obukhov49,Corrsin51}. Let $\theta$ be a homogeneous isotropic scalar field subjected to incompressible homogeneous isotropic turbulence. An extension of Kolmogorov's analysis \cite{K41a} to scalar turbulence yields
\begin{equation}
\langle[\theta(\mathbf{x+r})-\theta(\mathbf{x})]^2\rangle \propto\bepsilon^{-1/3} \bchi r^{2/3}
\end{equation}
for spatial separation $r$ within the inertial-convective range, where $\bepsilon$ and $\bchi$ are the mean dissipation rates of energy and scalar variance. An equivalent statement may be made in the Fourier space; scalar variance spectrum $E_\theta(k)$ satisfying $\frac{1}{2}\langle \theta^2\rangle$ $=$ $\int_0^\infty E_\theta(k) dk$ may take a universal -5/3 power law -- the \emph{Obukhov-Corrsin spectrum} -- within the inertial-convective range:
\begin{equation}
E_\theta(k) = K_\theta \bepsilon^{-1/3}\bchi k^{-5/3},
\label{OC spectrum}
\end{equation}
where $K_\theta$ is often referred to as the Obukhov-Corrsin constant. Since its discovery by Refs. \cite{Obukhov49,Corrsin51}, a number of works have examined and supported its universality from both experimental and numerical aspects \cite{Sreeni96, MW98, WCB99, YXS02, WG04, GW15}.

Besides the primitive knowledge of the passive-scalar turbulence based on a dimensional analysis, more profound understanding can be reached via 
dynamical equations of the scalar statistics. Second-order moment closure may be one successful category of such attempts, which enables quantitative description of the energy and scalar variance spectra on the basis of their dynamical modeling \cite{Zhou21}. The eddy-damped quasi-normal Markovian (EDQNM) model may be a typical example realizing simple spectral closure on scalar turbulence with the help of ad hoc eddy-damping terms \cite{Briard-etal16,Orszag70}. On the other hand, there are more self-consistent approaches based on the nature of the exact governing laws; abridged Lagrangian-history direct-interaction approximation (ALHDIA \cite{Kraichnan65,Kraichnan66b}) and Lagrangian-renormalization approximation (LRA \cite{Kaneda81,Kaneda86}) theories enable systematic derivations of the closure models of the scalar variance without relying on any empirical parameters (There is another approach by Ref. \cite{Goto-Kida99} combining Kraichnan's perturbation method \cite{Kraichnan59} and Kaneda's Lagrangian treatment, which finally reduces to a rederivation of LRA model). Due to their Lagrangian formalism properly removing the sweeping effect, both models successfully depicted the scale-local transfer of the scalar variance, reproducing the $-5/3$-power law of Eq. (\ref{OC spectrum}). Then the Lagrangian formalism should be firmly recognized as a key ingredient for the self-consistent closure of passive-scalar turbulence (There is an interesting attempt to extract Lagrangian time scale from EDQNM on scalar flux, which partially incorporates the Lagrangian framework into classical EDQNM \cite{Bos-Bertoglio06}).  For more comprehensive review of self-consistent turbulence closure, the authors refer the readers to Secs. III-IV of Ref. \cite{Zhou21} where the significance of Lagrangian picture is well summarized with the history of theoretical developments. \\

% (There is another methodology combining Krarichnan's perturbation method \cite{Kraichnan59} and Kaneda's Lagrangian position function, which rederives LRA closure model \cite{Kida-Goto97}.)
%There is also an interesting attempt to extract Lagrangian time scale from passive-scalar diffusion within the EDQNM framework, which results in reasonable prediction of velocity spectrum \cite{Bos-Bertoglio06}.

In spite of these remarkable successes, Lagrangian closure applied to passive scalar turbulence often suffers from its deficiency arising from the nature of the passive scalar itself; unlike the vectors and tensors, the scalar field does not reflect distortion of fluid elements, which makes the memory-fading effect caused by fluid's random straining indescribable. To see a rough sketch of the problem, we consider the governing equation of the passive scalar $\theta$:
\begin{equation}
(\partial_t+u_i\partial_i)\theta=\kappa \Delta \theta,
\label{theta-phys}
\end{equation}
where $\mathbf{u}$ is the fluid velocity field and $\kappa$ is the diffusion coefficient. In this paper we take summation for repeated indices of tensors and vectors. Now the Lagrangian derivative of $\theta$ may vanish in the inertial-convective range (essentially equivalent to a limit case $\kappa\to 0$), so the scalar value may be conserved along the Lagrangian trajectory. In the framework of turbulence closure, this is cast into the long-time memory effect of the scalar correlation. As a result, in the Lagrangian closure based on the scalar field, the memory of scalar field survives for long time exceeding the turbulence timescale, which indeed disagree to the Kolmogorov-Obukhov-Corrsin theory. Then, eventually, conventional Lagrangian closure overestimates the turbulence-mixing effect of scalar, lacking a quantitative predictability in scalar-variance spectrum. As carefully remarked by Refs. \cite{Kaneda81,Kaneda07}, turbulence closure may be improved by alternative choice of the representative variables properly representing the physics of our concern, which is a central issue to be adressed in the present work. A pioneering work of the \emph{rotation-invariant strain-based LRA} (RI-LRA \cite{GNK00}) chooses the pure-strain statistics as alternative representatives to conventional Lagrangian velocity statistics, which is essentially inspired by the idea of strain-based ALHDIA (SBALHDIA) \cite{KH78}. Application of RI-LRA to two-dimensional passive scalar turbulence yields, with the help of DNS data and some additional approximations, better prediction of the scalar-variance spectrum in the viscous-convective range, which is substantially due to consideration of the random straining motion of fluid. Likewise, modification on the representatives still has some possibilities to improve predictability of the scalar-closure models.

This paper provides a self-consistent closure theory of 3D passive scalar turbulence on the basis of LRA. In contrast to RI-LRA focusing on the velocity statistics, we modify the representatives of scalar field; instead of the scalar field itself, we treat its second-order derivative which effectively characterizes local scalar distribution via \emph{Hessian matrix}. Unlike scalar itself, its Hessian matrix can reflect fluid's turbulent motion, realizing the memory fading effect of the scalar statistics. Then we reach an alternative theory -- hereafter referred to as \emph{Hessian-based LRA} or simply \emph{HBLRA} -- which offers a self-consistent closure model for scalar statistics. Based on the Hessian of two-rank, its two-point statistics become eventually four-rank correlation tensors. Whereas constructed from such complex quantities, the resultant closure equations are simple and feasible enough in the actual calculation. A prototypical idea is briefly applied to inertial-particle problem in our recent work \cite{AYMY18}, while more comprehensive and generalized discussions may be given in the present paper focusing on passive scalar turbulence. Finally, as the first step among various possible applications, we apply HBLRA to the Obukhov-Corrsin spectrum and present its analytical solution accompanied by universal memory-fading functions of the scalar. In Sec. \ref{FORMULATION} we provide the general formulation of HBLRA. In Sec. \ref{INERTIAL CONVECTIVE},  the Obukhov-Corrsin spectrum of the scalar-variance spectrum is derived by applying HBLRA to inertial-convective range of well-developed turbulence, where the universal Obukhov-Corrsin constant is theoretically derived by full analysis of HBLRA equations.

%Kolmogorov's observation in the inertial-range scaling extends to the Kolmogorov-Obukhov-Corrsin scaling of passive scalar turbulence, and 

\section{Formulation}\label{FORMULATION}
In the present study, we focus our attention on homogeneous turbulence in three dimensional Euclidean space, so Fourier analysis offers a convenient platform for the forthcoming discussions. We utilize the Fourier transformation defined by the following integral operation $\mathcal{F}|^k_x\times$;
\begin{equation}
\mathcal{F}|^k_x\  \times = (2\pi)^{-3}\int d^3 x \exp[-i\mathbf{k}\cdot\mathbf{x}] \ \times
\end{equation}
which provide a one-to-one mapping from an arbitrary field function $f(\mathbf{x})$ to the corresponding Fourier spectrum $f(\mathbf{k})$ (the same main symbol employed for simple notation); i.e. $f(\mathbf{k})$$=$$\mathcal{F}|^k_x f(\mathbf{x})$.

\subsection{Scalar field}\label{SCALAR FIELD}
Applying the Fourier transformation to Eq. (\ref{theta-phys}) yields the scalar dynamical equation in the Fourier space;
\begin{equation}
\left(\partial_t+\kappa k^2\right)\theta(\mathbf{k},t)
=\frac{1}{i}k_a[\mathbf{k;p,q}]u_a(\mathbf{p},t)\theta(\mathbf{q},t),
\label{theta0}
\end{equation}
where $[\mathbf{k;p,q}]\equiv\iint d^3p\,d^3q\,\delta^3(\mathbf{k-p-q})\times$ represents a convolution in the wavenumber space. Then, in the Eulerian picture, the scalar distribution is developed by the bilinear coupling between velocity and scalar. Following Ref. \cite{Kaneda81}, the Lagrangian picture is introduced using the Lagrangian position function $\psi(\mathbf{x}',t;\mathbf{x},t')$ governed by $\partial_t \psi(\mathbf{x}',t;\mathbf{x},t')$ $+$ $\partial'_j[ u_j(\mathbf{x}',t)$ $\psi(\mathbf{x}',t;\mathbf{x},t')]$ $=$ $0$ and its initial condition $\psi(\mathbf{x},t';\mathbf{x}',t')$ $=$ $\delta^3(\mathbf{x}-\mathbf{x}')$ (see appendix \ref{POSITION FUNCTION}). Its Fourier space component $\psi(\mathbf{k}'',t;\mathbf{k},t')$ obeys
\begin{equation}
\partial_t\psi(\mathbf{k}'',t;\mathbf{k},t')
=ik''_b[\mathbf{k}'';\mathbf{-p,q}]u_b(\mathbf{p},t)\psi(\mathbf{q},t;\mathbf{k},t'),
\label{psi}
\end{equation}
with $\psi(\mathbf{k}',t';\mathbf{k},t')=\delta^3(\mathbf{k}'-\mathbf{k})$ (in this paper we define $\psi(\mathbf{k}',t;\mathbf{k},t')\equiv (2\pi)^{3}\mathcal{F}{}^{-k'}_{x'}|{}^k_x \psi(\mathbf{x}',t;\mathbf{x},t')$). In the Fourier representation, the Lagrangian scalar field reads:
\begin{equation}
\theta(\mathbf{k},t'|t)=\int \psi(\mathbf{k}'',t;\mathbf{k},t') \theta(\mathbf{k}'',t)\,d^3k'',
\end{equation}
which is governed by
\begin{equation}
\partial_t \theta(\mathbf{k},t'|t)=O(\kappa),
\label{L-scalar}
\end{equation}
where the bilinear coupling in the Eulerian equation is now lost. In case of sufficiently high P\'{e}clet number where $\kappa$ in Eq. (\ref{L-scalar}) becomes substantially negligible, Lagrangian scalar would keep its value, representing a long-time memory (absence of the memory-fading effect). Due to its invariance under arbitrary frame transformations, scalar field does not reflect the distortion and rotation of fluid elememt, but only the translation is cast into the scalar-field dynamics. In order to capture the memory fading due to straining motion, we need to extract some nature of scalar field that varies under the frame transformation.

\subsection{Hessian field}\label{HESSIAN FIELD}
When focusing on spatial distribution in finite local domains, one may observe some nontrivial structures of scalar field caused by turbulence mixing. Such a distribution has the directional characteristics dependent on frame transformation thus can be utilized in our closure strategy. Local distribution of the scalar may be typically characterized by its derivatives. A possibility of closure based on the scalar gradient has been discussed by Ref. \cite{GNK00}, where the distorting motion of the fluid can be reflected in the Lagrangian evolution of the scalar gradient. Unfortunately this variable choice yields insufficient memory-fading effect and thus not suitable to our objective. Then, other than the gradient vector, Hessian -- defined by $\partial_i\partial_j\theta$ in physical space -- may be the simplest and reasonable variable minimally characterizing the structure of the local scalar distribution. In addition, Hessian is capable of characterizing local maximum, minimum, and saddle points of the scalar distribution which could provide us more physically relevant properties compared with the scalar field itself. One thing should be noted, however, is that Hessian may be too much sensitive to small-scale structures for its derivative operations. Thence, in this paper, we deal with its non-local expression $\Delta^{-1}\partial_i\partial_j\theta$ to focus our attention on wider scale range, where $\Delta^{-1}$ is the inverse operator of Lapracian.
%\begin{equation*}
%\Delta^{-1}\equiv \iiint d^3 x'\frac{1}{4\pi |\mathbf{x}'-\mathbf{x}|}\ \times\ .
%\end{equation*}
In the Fourier space, this amounts to normalize the Hessian spectrum $k_ik_j\theta(\mathbf{k},t)$ by the wave number: 
\begin{equation}
\mathscr{H}^\mathrm{(p)}_{ij}(\mathbf{k},t)=\frac{k_ik_j}{k^2}\theta(\mathbf{k},t),
\end{equation}
which we term \emph{primary Hessian} for its later generalization. The governing equation of the primary Hessian is simply derived from Eq. (\ref{theta0}):
\begin{equation}
\begin{split}
\left(\partial_t+\kappa k^2\right)\mathscr{H}^\mathrm{(p)}_{ij}(\mathbf{k},t)
=&\frac{1}{i}k_a[\mathbf{k;p,q}]u_a(\mathbf{p},t)\mathscr{H}^\mathrm{(p)}_{ij}(\mathbf{q},t)\\
&+\frac{1}{i}k_a[\mathbf{k;p,q}]
X_{ij}(\mathbf{k},\mathbf{q})u_a(\mathbf{p},t)\theta(\mathbf{q},t),
\label{E-Hessian}
\end{split}
\end{equation}
where we split the right side into convection term (the first term) and the rest. Note that a geometrical factor $X_{ij}(\mathbf{k},\mathbf{q})\equiv k_ik_j/k^2-q_iq_j/q^2$
%\begin{equation}
%X_{ij}(\mathbf{k},\mathbf{q})
%=\frac{k_ik_j}{k^2}-\frac{q_iq_j}{q^2}
%\label{X vertex}
%\end{equation}
is traceless ($X_{ii}=0$), which guarantees the trace part of Eq. (\ref{E-Hessian}) to be the Eulerian-scalar equation (\ref{theta0}). Following the conventional LRA procedure, we shall introduce the Lagrangian variable of the Hessian:
\begin{equation}
\mathscr{H}^\mathrm{(p)}_{ij}(\mathbf{k},t'|t)
=\int \psi(\mathbf{k}'',t;\mathbf{k},t') \mathscr{H}^\mathrm{(p)}_{ij}(\mathbf{k}'',t) d^3k''
\end{equation}
which obeys
\begin{equation}
\partial_t \mathscr{H}^\mathrm{(p)}_{ij}(\mathbf{k},t'|t)
=\int d^3k'' \psi(\mathbf{k}'',t;\mathbf{k},t')
\left\{-\kappa k''{}^2 \mathscr{H}^\mathrm{(p)}_{ij}(\mathbf{k}'',t)
+\frac{1}{i}k''_a[\mathbf{k}'';\mathbf{p,q}]X_{ij}(\mathbf{k}'',\mathbf{q}) u_a(\mathbf{p},t)\theta(\mathbf{q},t)\right\}.
\label{L-Hessian-1}
\end{equation}
Here we should focus on the trace part of Eq. (\ref{L-Hessian-1}) reproducing the Lagrangian-scalar equation (\ref{L-scalar}). In other words, the trace part of the Lagrangian Hessian is insensitive to the bilinear coupling between $\mathbf{u}$ and $\theta$. This enables a natural extension of our Hessian $\mathscr{H}_{ij}$ in its trace part; using a real number $\xi$, we generalize our Hessian field as
\begin{equation}
\mathscr{H}_{ij}(\mathbf{k},t)
\equiv T_{ij}(\mathbf{k})\theta(\mathbf{k},t),\ \ \ 
T_{ij}(\mathbf{k})=\frac{k_ik_j}{k^2}-\xi\delta_{ij}.
\label{generalized H}
\end{equation}
As remarked in Ref. \cite{Kaneda81}, so called moment-closure approximation in non-linear systems demands suitable choices of the representative variables with carefully considering their physical meanings, which will be conducted for passive scalar turbulence in this study. The physical significance of the generalized Hessian $\mathscr{H}_{ij}$ depends on $\xi$; in addition to the primary Hessian ($\xi=0$), one may define solenoidal ($\xi=1$) and traceless Hessians($\xi=1/3$)  as physically relevant examples:
\begin{equation}
\mathscr{H}_{ij}(\mathbf{k},t)=
\begin{dcases}
\mathscr{H}^\mathrm{(s)}_{ij}(\mathbf{k},t)=&\left(\frac{k_ik_j}{k^2}-\delta_{ij}\right)\theta(\mathbf{k},t)\ \ (\textrm{solenoidal}),\\
\mathscr{H}^\mathrm{(t)}_{ij}(\mathbf{k},t)=&\left(\frac{k_ik_j}{k^2}-\frac{1}{3}\delta_{ij}\right)\theta(\mathbf{k},t)\ \ (\textrm{traceless}).
\end{dcases}
\end{equation}
The primary Hessian $\mathscr{H}^\mathrm{(p)}_{ij}$ corresponds to a non-local expression in physical space, i.e. $\Delta^{-1}\partial_i\partial_j \theta$, which still characterizes local structure of the scalar distribution; spherical spots may be detected by its trace part while sheets or filaments by the rest deviatric components. The traceless Hessian, the deviatric part of the primary Hessian, then emphasizes sheet-like or filament-like structures while diminishes in spherical spots or voids. Also, unlike the primary Hessian containing $\theta$ in its trace, the traceless Hessian is completely free from long-time memory of $\theta$, which may leads to the shortest correlation time scale among others. The solenoidal Hessian represents divergence-free part of the primary Hessian whose divergence exactly coincides with the scalar gradient; $\partial_j\Delta^{-1}\partial_i\partial_j\theta = \partial_i\theta$. Thence the solenoidal Hessian cancels out the contribution from the scalar gradient. Providing there is any characteristic structure of scalar (spot, sheet, filament, etc.) surrounded by some larger-scale structure, primary and traceless Hessians may be affected by this larger structure acting like a background gradient. Thus solenoidal Hessian may be suitable to quantify local scalar structures without interference from such larger-scale structures. Lagrangian variable of the generalized Hessian $\mathscr{H}_{ij}$ (hereafter we simply call it Hessian) obeys
\begin{equation}
\partial_t \mathscr{H}_{ij}(\mathbf{k},t'|t)
=\int d^3k'' \psi(\mathbf{k}'',t;\mathbf{k},t')
\left\{-\kappa k''{}^2 \mathscr{H}_{ij}(\mathbf{k}'',t)
+\frac{1}{i}k''_a[\mathbf{k}'';\mathbf{p,q}]X_{ij}(\mathbf{k}'',\mathbf{q}) u_a(\mathbf{p},t)\theta(\mathbf{q},t)\right\}
\label{L-Hessian0}
\end{equation}
which is identical to Eq. (\ref{L-Hessian-1}). 
%Note that $\xi$ should be chosen so that its corresponding Hessian represents certain significance in the physical process we focus on.

In what follows we should describe all the dynamical equations in terms of Hessian regarded as a principal dynamical variable, for which we need to express $\theta(\mathbf{k},t)$ in terms of $\mathscr{H}_{ij}(\mathbf{k},t)$. Then we introduce a projection operator, say $\mathsfbi{Z}$, from the Hessian-valued to scalar-valued functions: $Z_{ij}(\mathbf{k})\,\mathscr{H}_{ij}(\mathbf{k},t)$ $=\theta(\mathbf{k},t)$, where 
\begin{equation}
Z_{ij}(\mathbf{k})\equiv \frac{T_{ij}(\mathbf{k})}{T^2}\ \   (T\equiv\lVert\mathsfbi{T}\rVert=\sqrt{T_{ab}T_{ab}})
\end{equation}
which satisfies $T_{ij}(\mathbf{k})Z_{ij}(\mathbf{k})=1$. Reminding $\mathsfbi{T}$ is related with $\xi$ by Eq. (\ref{generalized H}), we have $T^2$$=$$3\xi^2-2\xi+1$ which takes values 1 (primary), 2/3 (traceless), and 2 (solenoidal). Now Eq. (\ref{L-Hessian0}) can be rewritten in terms of Hessian alone:
\begin{equation}
\begin{split}
&\partial_t \mathscr{H}_{ij}(\mathbf{k},t'|t)\\
&=\int d^3k'' \psi(\mathbf{k}'',t;\mathbf{k},t')
\left\{-\kappa k''{}^2 \mathscr{H}_{ij}(\mathbf{k}'',t)
+[\mathbf{k}'';\mathbf{p,q}]Y_{ij.a.bc}(\mathbf{k}'',\mathbf{q}) u_a(\mathbf{p},t)\mathscr{H}_{bc}(\mathbf{q},t)\right\},
\end{split}
\label{L-Hessian}
\end{equation}
where $Y_{ij.a.bc}(\mathbf{k}'',\mathbf{q})=X_{ij}(\mathbf{k}'',\mathbf{q})k''_a Z_{bc}(\mathbf{q})/i$.

%\begin{equation}
%\theta(\mathbf{k},t)=Z_{ij}(\mathbf{k})H^{\scalebox{0.6}{n}}_{ij}(\mathbf{k},t),\ 
%Z_{ij}(\mathbf{k})=\beta\left(\alpha\frac{k_ik_j}{k^2}+(1-\alpha)\delta_{ij}\right)
%\label{decoding 1}
%\end{equation}

%where $\alpha,\beta(\in\mathbb{R})$ is to be determined so that $Z_{ij}(\mathbf{k})T^{\scalebox{0.6}{n}}_{ij}(\mathbf{k})=1$. For instance we obtain 
%\begin{subequations}
%\begin{align}
%Z^{{}_0}_{ij}(\mathbf{k})&=\alpha\frac{k_ik_j}{k^2}+(1-\alpha)\delta_{ij}\ (\beta=1),\\
%Z^{{}_{1/3}}_{ij}(\mathbf{k})&=\frac{3}{2}\frac{k_ik_j}{k^2}+\left(\beta-\frac{3}{2}\right)\delta_{ij}\ (\alpha\beta=3/2).
%\end{align}
%\end{subequations}
%In particular case of $n=1/3$ (traceless), isotropic part $(\beta-\frac{3}{2})\delta_{ij}$ does not contribute at all, so we can choose $(\alpha,\beta)=(1,3/2)$ yielding $Z^{{}_{1/3}}_{ij}(\mathbf{k})=\frac{3}{2}\frac{k_ik_j}{k^2}$. 

\subsection{Lagrangian response}
Following LRA procedure, we shall introduce the response of $\mathscr{H}_{ij}(\mathbf{k},t'|t)$ against infinitesimal disturbance. For this sake, we consider an infinitesimal disturbance $\delta f_\theta$ on the scalar field $\theta$ of Eq. (\ref{theta0}):
\begin{equation}
\left(\partial_t+\kappa k^2\right)\theta(\mathbf{k},t)
=\frac{1}{i}k_a[\mathbf{k;p,q}]u_a(\mathbf{p},t)\theta(\mathbf{q},t)
+\delta f_\theta(\mathbf{k},t')\delta(t-t').
\end{equation}
Multiplying this by $T_{ij}(\mathbf{k})$ yields a disturbed Eulerian Hessian equation:
\begin{equation}
\begin{split}
\left(\partial_t+\kappa k^2\right)\mathscr{H}_{ij}(\mathbf{k},t)
=&\frac{1}{i}k_a[\mathbf{k;p,q}]u_a(\mathbf{p},t)\mathscr{H}_{ij}(\mathbf{q},t)
+[\mathbf{k;p,q}]Y_{ij.a.mn}(\mathbf{k},\mathbf{q})
u_a(\mathbf{p},t)\mathscr{H}_{mn}(\mathbf{q},t)\\
&+T_{ij}(\mathbf{k})\delta f_\theta(\mathbf{k},t')\delta(t-t').
\end{split}
\end{equation}
The infinitesimal variation $\delta \mathscr{H}_{ij}(\mathbf{k},t)$ may be written as a linear functional of $\delta f_\theta(\mathbf{k}',t')$, which may be expressed by the functional derivative. We further rewrite this as
\begin{equation}
\delta \mathscr{H}_{ij}(\mathbf{k},t')
=\int d^3k'\frac{\delta \mathscr{H}_{ij}(\mathbf{k},t)}{\delta f_\theta(\mathbf{k}',t')} \delta f_\theta(\mathbf{k}',t')
=\int d^3k'\frac{\delta \mathscr{H}_{ij}(\mathbf{k},t)}{\delta f_\theta(\mathbf{k}',t')} Z_{lm}(\mathbf{k}') T_{lm}(\mathbf{k}')\delta f_\theta(\mathbf{k}',t'),
\end{equation}
where $\delta/\delta f_\theta(\mathbf{k}',t')$ is functional derivative operation. Regarding $\delta F_{ij}(\mathbf{k},t)\equiv T_{ij}(\mathbf{k}) \delta f_\theta(\mathbf{k},t)$ as the disturbance tensor applied to $\mathscr{H}_{ij}(\mathbf{k},t)$, the Eulerian response function of the Hessian may be
\begin{equation}
\mathscr{G}^{\scalebox{0.6}{E}}_{ij.lm}(\mathbf{k},t;\mathbf{k}',t')
=\frac{\delta \mathscr{H}_{ij}(\mathbf{k},t)}{\delta f_\theta(\mathbf{k}',t')} Z_{lm}(\mathbf{k}').
\end{equation}
In the same manner, the Lagrangian response function reads
\begin{equation}
\mathscr{G}^{\scalebox{0.6}{L}}_{ij.lm}(\mathbf{k},t;\mathbf{k}',t')
=\frac{\delta \mathscr{H}_{ij}(\mathbf{k},t'|t)}{\delta f_\theta(\mathbf{k}',t')} Z_{lm}(\mathbf{k}').
\end{equation}
Functional derivative on Eq. (\ref{L-Hessian}) yields the equation for the Lagrangian response function $\mathscr{G}^{\scalebox{0.6}{L}}_{ij.lm}(\mathbf{k},t;\mathbf{k}',t')$:
\begin{equation}
\begin{split}
\partial_t \mathscr{G}^{\scalebox{0.6}{L}}_{ij.lm}(\mathbf{k},t;\mathbf{k}',t')
&=\frac{\delta}{\delta f_\theta(\mathbf{k}',t')}
\partial_t \mathscr{H}_{ij}(\mathbf{k},t'|t)Z_{lm}(\mathbf{k}')\\
&=\int d^3k'' \psi(\mathbf{k}'',t;\mathbf{k},t')
[\mathbf{k}'';\mathbf{p,q}]Y_{ij.a.bc}(\mathbf{k}'',\mathbf{q}) u_a(\mathbf{p},t)\frac{\delta \mathscr{H}_{bc}(\mathbf{q},t)}{\delta f_\theta(\mathbf{k}',t')}Z_{lm}(\mathbf{k}')\\
&\ \ \ \ -\int d^3k'' \psi(\mathbf{k}'',t;\mathbf{k},t') \kappa k''{}^2 
\frac{\delta \mathscr{H}_{ij}(\mathbf{k}'',t)}{\delta f_\theta(\mathbf{k}',t')}Z_{lm}(\mathbf{k}')\\
&=\int d^3k'' \psi(\mathbf{k}'',t;\mathbf{k},t')
[\mathbf{k}'';\mathbf{p,q}]Y_{ij.a.bc}(\mathbf{k}'',\mathbf{q}) u_a(\mathbf{p},t)\mathscr{G}^{\scalebox{0.6}{E}}_{bc.lm}(\mathbf{q},t;\mathbf{k}',t')\\
&\ \ \ \ -\int d^3k'' \psi(\mathbf{k}'',t;\mathbf{k},t') \kappa k''{}^2 
\mathscr{G}^{\scalebox{0.6}{E}}_{ij.lm}(\mathbf{k}'',t;\mathbf{k}',t')
\end{split}
\label{L-G eq.}
\end{equation}
for $t\geq t'$, which is accompanied by its initial condition:
\begin{equation}
\mathscr{G}^{\scalebox{0.6}{L}}_{ij.lm}(\mathbf{k},t';\mathbf{k}',t')=
\mathscr{G}^{\scalebox{0.6}{E}}_{ij.lm}(\mathbf{k},t';\mathbf{k}',t')=
T_{ij}(\mathbf{k})\delta^3(\mathbf{k}-\mathbf{k}')Z_{lm}(\mathbf{k}').
\end{equation}

\subsection{Perturbation analysis}
Here we review the dynamical equations which our LRA procedures are based on. Then our system is totally described by the following equations (Eq. (\ref{EH eq.}) below may be soon reached by rewriting $\theta$ in the right-side of Eq. (\ref{theta0}) into $\mathsfbi{Z}\cdot\boldsymbol{\mathscr{H}}$ and multiplying both sides by $\mathsfbi{T}$.):
\begin{equation}
\left(\partial_t+\kappa k^2\right)\mathscr{H}^{\scalebox{0.6}{E}}_{ij}(\mathbf{k},t)
=\lambda\frac{1}{i}T_{ij}(\mathbf{k})k_c[\mathbf{k;p,q}]Z_{ab}(\mathbf{q})u_c(\mathbf{p},t)\mathscr{H}^{\scalebox{0.6}{E}}_{ab}(\mathbf{q},t),
\label{EH eq.}
\end{equation}
\begin{equation}
\left(\partial_t+\kappa k^2\right)\mathscr{G}^{\scalebox{0.6}{E}}_{ij.lm}(\mathbf{k},t;\mathbf{k}',t')
=\lambda\frac{1}{i}T_{ij}(\mathbf{k})k_c[\mathbf{k;p,q}]Z_{ab}(\mathbf{q})u_c(\mathbf{p},t)\mathscr{G}^{\scalebox{0.6}{E}}_{ab.lm}(\mathbf{k},t;\mathbf{k}',t')\ \ (t\geq t'),
\label{EG eq.}
\end{equation}
\begin{equation}
\begin{split}
\partial_t\mathscr{H}^{\scalebox{0.6}{L}}_{ij}(\mathbf{k},t'|t)
=&\lambda\int d^3k'' \psi(\mathbf{k}'',t;\mathbf{k},t')
[\mathbf{k}'';\mathbf{p,q}]Y_{ij.a.bc}(\mathbf{k}'',\mathbf{q}) u_a(\mathbf{p},t)\mathscr{H}^{\scalebox{0.6}{E}}_{bc}(\mathbf{q},t)\\
&-\kappa\int d^3k'' \psi(\mathbf{k}'',t;\mathbf{k},t') k''{}^2 
\mathscr{H}^{\scalebox{0.6}{E}}_{ij}(\mathbf{k}'',t),
\end{split}
\label{LH eq.}
\end{equation}
\begin{equation}
\begin{split}
\partial_t\mathscr{G}^{\scalebox{0.6}{L}}_{ij.lm}(\mathbf{k},t;\mathbf{k}',t')
&=\lambda\int d^3k'' \psi(\mathbf{k}'',t;\mathbf{k},t')
[\mathbf{k}'';\mathbf{p,q}]Y_{ij.a.bc}(\mathbf{k}'',\mathbf{q}) u_a(\mathbf{p},t)\mathscr{G}^{\scalebox{0.6}{E}}_{bc.lm}(\mathbf{q},t;\mathbf{k}',t')\\
&\ \ \ \ -\kappa\int d^3k'' \psi(\mathbf{k}'',t;\mathbf{k},t') k''{}^2 
\mathscr{G}^{\scalebox{0.6}{E}}_{ij.lm}(\mathbf{k}'',t;\mathbf{k}',t')\ \ (t\geq t'), 
\end{split}
\label{LG eq.}
\end{equation}
\begin{equation}
\partial_t\psi(\mathbf{k}'',t;\mathbf{k},t')
=\lambda ik''_b[\mathbf{k}'';\mathbf{-p,q}]u_b(\mathbf{p},t)\psi(\mathbf{q},t;\mathbf{k},t'),
\label{psi 1}
\end{equation}
\begin{equation}
\mathscr{G}^{\scalebox{0.6}{L}}_{ij.lm}(\mathbf{k},t';\mathbf{k}',t')=
\mathscr{G}^{\scalebox{0.6}{E}}_{ij.lm}(\mathbf{k},t';\mathbf{k}',t')=
T_{ij}(\mathbf{k})\delta^3(\mathbf{k}-\mathbf{k}')Z_{lm}(\mathbf{k}'),
\label{initial G}
\end{equation}
\begin{equation}
\psi(\mathbf{k}'',t';\mathbf{k},t')=\delta^3(\mathbf{k}''-\mathbf{k}),
\label{initial P}
\end{equation}
where we attach the subscripts ``L" and ``E" on the Lagrangian and Eulerian $\mathscr{H}_{ij}$ for clarity of notations. Also we attach $\lambda (=1)$ as a bookkeeping parameter for later perturbation analysis. Then we define non-perturbative dynamics by $O(\lambda^0)$ analysis of Eqs. (\ref{EH eq.})-(\ref{initial P}):
\begin{equation}
\left(\partial_t+\kappa k^2\right)\tilde{\mathscr{H}}_{ij}(\mathbf{k},t)=0,
\end{equation}
\begin{equation}
\left(\partial_t+\kappa k^2\right)\tilde{\mathscr{G}}_{ij.lm}(\mathbf{k},t;\mathbf{k}',t')=0,\ (t\geq t')
\end{equation}
\begin{equation}
\tilde{\mathscr{G}}_{ij.lm}(\mathbf{k},t';\mathbf{k}',t')=
T_{ij}(\mathbf{k})\delta^3(\mathbf{k}-\mathbf{k}')Z_{lm}(\mathbf{k}'),
\end{equation}
\begin{equation}
\partial_t\tilde{\psi}(\mathbf{k}'',t;\mathbf{k},t')=0,
\end{equation}
where we attach tilde on the main symbols of the original fields. Also note that Eulerian and Lagrangian equations reduce to identical forms at this stage. In exactly the same manner, we shall introduce non-perturbative velocity $\tilde{u}_i(\mathbf{k},t)$ and response $\tilde{G}_{ij}(\mathbf{k},t;\mathbf{k}',t')$ (see Appendix \ref{EULER-LAGRANGE VELOCITY}). Then, using iterative time integration, both Lagrangian and Eulerian quantities can be expanded in terms of non-perturbative quantities. As a consequence, all the statistical quantities are expanded by the non-perturbative statistics; for the non-perturbative Hessian field we have
\begin{equation}
\tilde{\mathcal{H}}_{ij.lm}(\mathbf{k},t;\mathbf{k}',t')
\equiv \langle \tilde{\mathscr{H}}_{ij}(\mathbf{k},t)\tilde{\mathscr{H}}_{lm}(\mathbf{k}',t')\rangle,
\label{np 2TC}
\end{equation}
\begin{equation}
\tilde{\mathcal{G}}_{ij.lm}(\mathbf{k},t;\mathbf{k}',t')
\equiv \langle \tilde{\mathscr{G}}_{ij.lm}(\mathbf{k},t;\mathbf{k}',t')\rangle
=\tilde{\mathscr{G}}_{ij.lm}(\mathbf{k},t;\mathbf{k}',t'),
\end{equation}
likewise for the velocity field $\tilde{Q}_{ij}(\mathbf{k},t;\mathbf{k}',t')$ $\equiv \langle \tilde{u}_i(\mathbf{k},t)\tilde{u}_j(\mathbf{k}',t')\rangle$ and $
\tilde{G}_{ij}(\mathbf{k},t;\mathbf{k}',t')$. Note that the velocity and the Hessian are to be uncorrelated with each other at the non-perturbative stage due to the absence of the non-linear coupling between them.

\if0
In the primitive perturbation analysis, all the dynamical fields can be expanded in terms of non-perturbative fields; e.g. Eulerian Hessian can be expanded as
\begin{equation}
\begin{split}
\mathscr{H}^{\scalebox{0.6}{E}}_{ij}(\mathbf{k},t)
=&\tilde{\mathscr{H}}_{ij}(\mathbf{k},t)\\
&+\lambda\int^t_{t_0} ds\ \int d^3k'\ \tilde{\mathscr{G}}_{ij.lm}(\mathbf{k},t;\mathbf{k}',s)
\frac{1}{i}T_{lm}(\mathbf{k}')k_c[\mathbf{k}';\mathbf{p,q}]Z_{ab}(\mathbf{q})\tilde{u}_c(\mathbf{p},s)\tilde{\mathscr{H}}_{ab}(\mathbf{q},s)\\
&+O(\lambda^2),
\end{split}
\end{equation} 
where $t_0$ is the initial time. 
\fi

\subsection{Lagrangian renormalization}\label{LAGRANGIAN RENORMALIZATION}
Now we are prepared to obtain the moment closure equations based on the renormalized perturbation analysis. Departing from simple perturbation, renormalized perturbation analysis enables to describe stochastic relaxation of the correlations, which is the key to successful closure model at general Reynolds and Schmidt numbers. The renormalization machinery enables a systematic derivation of the closed system for the two-time correlations and the averaged responses:
\begin{equation}
\begin{split}
\mathcal{H}_{ij.lm}(\mathbf{k},t;\mathbf{k}',t')
&\equiv\langle \mathscr{H}^{\scalebox{0.6}{L}}_{ij}(\mathbf{k},t'|t)\mathscr{H}^{\scalebox{0.6}{E}}_{lm}(\mathbf{k}',t')\rangle,\\
\mathcal{G}_{ij.lm}(\mathbf{k},t;\mathbf{k}',t')
&\equiv \langle \mathscr{G}^{\scalebox{0.6}{L}}_{ij.lm}(\mathbf{k},t;\mathbf{k}',t')\rangle\ \ (t \geq t'),\\
Q_{ij}(\mathbf{k},t;\mathbf{k}',t')
&\equiv P_{ia}(\mathbf{k})\langle u^{\scalebox{0.6}{L}}_a(\mathbf{k},t'|t)u^{\scalebox{0.6}{E}}_j
(\mathbf{k}',t')\rangle\ \ (t \geq t'),\\
G_{ij}(\mathbf{k},t;\mathbf{k}',t')
&\equiv P_{ia}(\mathbf{k})\langle G^{\scalebox{0.6}{L}}_{ab}(\mathbf{k},t;\mathbf{k}',t')\rangle P_{bj}(\mathbf{k}')\ \ (t \geq t'),
\end{split}
\label{renormalized quantities}
\end{equation}
where $P_{ij}(\mathbf{k})$ ($\equiv \delta_{ij}-k_ik_j/k^2$) is the solenoidal operator. Due to homogeneity, these are further simplified as $\mathcal{H}_{ij.lm}(\mathbf{k},t;\mathbf{k}',t')=\delta^3(\mathbf{k}+\mathbf{k}')$ $\mathcal{H}_{ij.lm}(\mathbf{k};t,t')$, $\mathcal{G}_{ij.lm}(\mathbf{k},t;\mathbf{k}',t')
=\delta^3(\mathbf{k}-\mathbf{k}')$ $\mathcal{G}_{ij.lm}(\mathbf{k};t,t')$, $Q_{ij}(\mathbf{k},t;\mathbf{k}',t')
=\delta^3(\mathbf{k}+\mathbf{k}')$ $Q_{ij}(\mathbf{k};t,t')$, $G_{ij}(\mathbf{k},t;\mathbf{k}',t')
=\delta^3(\mathbf{k}+\mathbf{k}')$ $G_{ij}(\mathbf{k};t,t')$. In the context of renormalization theories, these statistical variables are treated as the \emph{renormalized} variables, while the non-perturbative counterpart recognized as the \emph{bare} ones. Renormalized quantities are obtained as partial summation of infinite perturbation series, where multiple steps of triad interactions are incorporated inside. Then, the renormalization may be rephrased as the replacement of the expansion basis from bare to renormalized ones so that the stochastic relaxation process can be considered even at finite order (for more details see appendix \ref{RENORMALIZATION}).\\

%After the first introduction by Ref. \citep{Wyld61}, the renormazliation procedure in turbulence theory have been reinterpreted and developed by several authors \citep{Kraichnan77,Kaneda81}. 

Applying renormalization to the moment equations of $\mathcal{H}_{ij.lm}(\mathbf{k},t;\mathbf{k}',t')$, $\mathcal{G}_{ij.lm}(\mathbf{k},t;\mathbf{k}',t')$,
$Q_{ij}(\mathbf{k},t;\mathbf{k}',t')$, and $G_{ij}(\mathbf{k},t;\mathbf{k}',t')$, these are consistently closed. Here we present the closure equations of the Hessian statistics;  
\begin{equation}
\begin{split}
&(\partial_t+2\kappa k^2)\mathcal{H}_{ij.lm}(\mathbf{k},t,t)\\
=&2T_{ij}(\mathbf{k})T_{lm}(\mathbf{k})
[\mathbf{k};\mathbf{p},\mathbf{q}]\int^t_{t_0}ds\, Q_{ab}(\mathbf{p};t,s)
\left\{k_ak_bH(\mathbf{q};s,t)G_H(-\mathbf{k};t,s)
-k_aq_b H(-\mathbf{k};s,t)G_H(\mathbf{q};t,s)\right\},
\end{split}
\label{ETC eq.0}
\end{equation}

\begin{equation}
\begin{split}
&(\partial_t+\kappa k^2)\mathcal{H}_{ij.lm}(\mathbf{k};t,t')\\
=&-[\mathbf{k};\mathbf{p},\mathbf{q}]
X_{ij}(\mathbf{k},\mathbf{q})T_{lm}(\mathbf{k})q_aq_b \int^t_{t'} ds\, Q_{ab}(\mathbf{p};t,s)H(-\mathbf{k},t,t')\\
&-[\mathbf{k};\mathbf{p},\mathbf{q}]X_{ij}(\mathbf{k},\mathbf{q})T_{lm}(\mathbf{k})k_aq_b\int^t_{t_0}ds\, 
Q_{ab}(\mathbf{p};t,s)G_H(\mathbf{q};t,s)H(-\mathbf{k};s,t')\\
&+[\mathbf{k};\mathbf{p},\mathbf{q}]X_{ij}(\mathbf{k},\mathbf{q})T_{lm}(\mathbf{k})
k_ak_b\int^{t'}_{t_0}ds\, 
Q_{ab}(\mathbf{p};t,s)H(\mathbf{q};t,s)G_H(-\mathbf{k};t',s),
\end{split}
\label{2TC eq.0}
\end{equation}

\begin{equation}
\begin{split}
&(\partial_t+\kappa k^2) \mathcal{G}_{ij.lm}(\mathbf{k};t,t')\\
=&-[\mathbf{k};\mathbf{p},\mathbf{q}]
X_{ij}(\mathbf{k},\mathbf{q})Z_{lm}(\mathbf{k})q_aq_b\int^t_{t'} ds\, Q_{ab}(\mathbf{p};t,s)G_H(\mathbf{k},t,t')\\
&-[\mathbf{k};\mathbf{p},\mathbf{q}]X_{ij}(\mathbf{k},\mathbf{q})Z_{lm}(\mathbf{k})k_aq_b\int^t_{t_0}ds\, 
Q_{ab}(\mathbf{p};t,s)G_H(\mathbf{q};t,s)G_H(\mathbf{k};s,t')\ (t\geq t'),
\end{split}
\label{G eq.0}
\end{equation}

\begin{equation}
\mathcal{G}_{ij.lm}(\mathbf{k};t',t')=T_{ij}(\mathbf{k})Z_{lm}(\mathbf{k}),
\label{G eq.0 initial}
\end{equation}
where $H(\mathbf{k};t,t')$ and $G_H(\mathbf{k};t,t')$ are projected components of $\mathcal{H}_{ij.lm}(\mathbf{k};t,t')$ and $\mathcal{G}_{ij.lm}(\mathbf{k};t,t')$:
\begin{equation}
\begin{split}
H(\mathbf{k};t,t')=Z_{ij}(\mathbf{k})\mathcal{H}_{ij.lm}(\mathbf{k};t,t')Z_{lm}(\mathbf{k}),
\end{split}
\label{scalar H}
\end{equation}
\begin{equation}
\begin{split}
G_H(\mathbf{k};t,t')
=Z_{ij}(\mathbf{k})\mathcal{G}_{ij.lm}(\mathbf{k};t,t')T_{lm}(\mathbf{k}).
\end{split}
\label{scalar G}
\end{equation}
Now Eqs. (\ref{ETC eq.0})-(\ref{G eq.0}) combined with velocity closure (equivalent to Eqs. (2.35)-(2.46) of Ref. \cite{Kaneda81}) form a closed set of equations for $\mathcal{H}$, $\mathcal{G}$, $Q$, and $G$ ($F$ in Ref. \cite{Kaneda81}). Among the total dynamical variables, Hessian statistics have totally twelve degrees of freedom. On the other hand, we soon realize that only $H(\mathbf{k};t,t')$ and $G_H(\mathbf{k};t,t')$ appear in the right sides of Eqs. (\ref{ETC eq.0})-(\ref{G eq.0}), suggesting only limited degrees of freedom among all the tensor components are relevant in the total dynamics. Indeed, Eq. (\ref{ETC eq.0}) $\times Z_{ij}Z_{lm}$, Eq. (\ref{2TC eq.0}) $\times Z_{ij}Z_{lm}$, Eq. (\ref{G eq.0}) $\times Z_{ij}T_{lm}$, and Eq. (\ref{G eq.0 initial}) $\times Z_{ij}T_{lm}$ lead to a closed set of equations for $H(\mathbf{k};t,t')$ and $G_H(\mathbf{k};t,t')$:
\begin{equation}
\begin{split}
&\left(\partial_t+2\kappa k^2\right)H(\mathbf{k};t,t)\\
&=2[\mathbf{k};\mathbf{p},\mathbf{q}]\int^t_{t_0}ds\, Q_{ab}(\mathbf{p};t,s)
\left\{k_ak_b H(\mathbf{q};s,t)G_H(-\mathbf{k};t,s)
-k_aq_b H(-\mathbf{k};s,t)G_H(\mathbf{q};t,s)\right\},
\end{split}
\label{ETC LRA'}
\end{equation}

\begin{equation}
\begin{split}
\left(\partial_t+\kappa k^2\right)&H(\mathbf{k};t,t')\\
=&-\frac{1}{T^2}\frac{k_ak_b}{k^2}[\mathbf{k};\mathbf{p},\mathbf{q}]
X_{ab}(\mathbf{k},\mathbf{q})q_cq_d \int^t_{t'} ds Q_{cd}(\mathbf{p};t,s) H(-\mathbf{k},t,t')\\
&-\frac{1}{T^2}\frac{k_ak_b}{k^2}[\mathbf{k};\mathbf{p},\mathbf{q}]X_{ab}(\mathbf{k},\mathbf{q})k_cq_d\int^t_{t_0}ds\, 
Q_{cd}(\mathbf{p};t,s)G_H(\mathbf{q};t,s) H(-\mathbf{k};s,t')\\
&+\frac{1}{T^2}\frac{k_ak_b}{k^2}[\mathbf{k};\mathbf{p},\mathbf{q}]X_{ab}(\mathbf{k},\mathbf{q})k_ck_d\int^{t'}_{t_0}ds\, 
Q_{cd}(\mathbf{p};t,s) H(\mathbf{q};t,s)G_H(-\mathbf{k};t',s),
\end{split}
\label{2TC LRA'}
\end{equation}

\begin{equation}
\begin{split}
\left(\partial_t+\kappa k^2\right)& G_H(\mathbf{k};t,t')\\
=&-\frac{1}{T^2}\frac{k_ak_b}{k^2}[\mathbf{k};\mathbf{p},\mathbf{q}]
X_{ab}(\mathbf{k},\mathbf{q})q_cq_d \int^t_{t'} ds Q_{cd}(\mathbf{p};t,s)G_H(\mathbf{k},t,t')\\
&-\frac{1}{T^2}\frac{k_ak_b}{k^2}[\mathbf{k};\mathbf{p},\mathbf{q}]X_{ab}(\mathbf{k},\mathbf{q})k_cq_d\int^t_{t'}ds\, 
Q_{cd}(\mathbf{p};t,s)G_H(\mathbf{q};t,s)G_H(\mathbf{k};s,t')\ (t\geq t'),
\end{split}
\label{G LRA'}
\end{equation}

\begin{equation}
G_H(\mathbf{k};t',t')=1.
\label{G LRA' initial}
\end{equation}
Note that a common factor $T^{-2}$ ($T$=$\lVert\mathsfbi{T}\rVert$) appears in every term on the right sides of Eqs. (\ref{2TC LRA'}) and (\ref{G LRA'}). Recalling that $\mathsfbi{T}$ determines representative variables, we find that the choice of representatives does not alter the structure of equations but determines the characteristic time scales of $H(\mathbf{k};t,t')$ and $G_H(\mathbf{k};t,t')$ via $T^{-2}$.

\subsection{Relation to scalar-based LRA}
The two-time correlation and response are to be decomposed into the trace and traceless parts:
\begin{equation}
\mathcal{H}_{ij.lm}(\mathbf{k};t,t')
=\left[\mathcal{A}_{ij}(\mathbf{k};t,t')+\left(\frac{1}{3}
-\xi\right)\delta_{ij}\Theta(\mathbf{k};t,t')\right]T_{lm}(\mathbf{k}),
\label{decomposed 2TC}
\end{equation}

\begin{equation}
\mathcal{G}_{ij.lm}(\mathbf{k};t,t')
=\left[\mathcal{B}_{ij}(\mathbf{k};t,t')+\left(\frac{1}{3}
-\xi\right)\delta_{ij}G_\theta(\mathbf{k};t,t')\right]Z_{lm}(\mathbf{k}),
\label{decomposed G}
\end{equation}
where $\mathcal{A}_{ij}$ and $\mathcal{B}_{ij}$ are traceless symmetric tensors; i.e. $\mathcal{A}_{jj}=\mathcal{B}_{jj}=0$. $\Theta$ and $G_\theta$ are the Lagrangian correlation and response of the scalar field which are chosen as the representative variables of the scalar-based LRA \cite{Kaneda86}. Then our representative variables $\mathcal{H}_{ij.lm}(\mathbf{k};t,t')$ and $\mathcal{G}_{ij.lm}(\mathbf{k};t,t')$ are endowed with ten degrees of freedom from $\mathcal{A}_{ij}$ and $\mathcal{B}_{ij}$ besides the two scalar functions $\Theta$ and $G_\theta$ of scalar-based LRA. Operations (\ref{scalar H}) and (\ref{scalar G}) yields
\begin{equation}
H(\mathbf{k};t,t')=\frac{1}{T^2}\frac{k_ik_j}{k^2}\mathcal{A}_{ij}(\mathbf{k};t,t')
+\left(1-\frac{2}{3T^2}\right)\Theta(\mathbf{k};t,t'),
\label{composite H}
\end{equation}
\begin{equation}
G_H(\mathbf{k};t,t')=\frac{1}{T^2}\frac{k_ik_j}{k^2}\mathcal{B}_{ij}(\mathbf{k};t,t')
+\left(1-\frac{2}{3T^2}\right)G_\theta(\mathbf{k};t,t').
\label{composite G}
\end{equation}
Thus our closure variables $H(\mathbf{k};t,t')$ and $G_H(\mathbf{k};t,t')$ are generalization of the Lagrangian scalar statistics $\Theta(\mathbf{k};t,t')$ and $G_\theta(\mathbf{k};t,t')$ by incorporating traceless components $\mathcal{A}_{ij}(\mathbf{k};t,t')$ and $\mathcal{B}_{ij}(\mathbf{k};t,t')$. Indeed the scalar-based LRA of Ref. \cite{Kaneda86} is exactly reproduced in the limit case $T^2\to\infty$, which is obvious from Eqs. (\ref{ETC LRA'})-(\ref{G LRA' initial}) and Eqs. (\ref{composite H})-(\ref{composite G}). 

\if0
Indeed, Eq. (\ref{2TC eq.0})$\times\delta_{ij}Z_{lm}$, Eq. (\ref{G eq.0})$\times\delta_{ij}T_{lm}$, and Eq. (\ref{G eq.0 initial})$\times\delta_{ij}T_{lm}$ read 
\begin{equation}
(\partial_t+\kappa k^2) \Theta(\mathbf{k};t,t')=0,
\label{Theta eq.}
\end{equation}
\begin{equation}
(\partial_t+\kappa k^2) G_\theta(\mathbf{k};t,t')=0,
\label{Gtheta eq.}
\end{equation}
\begin{equation}
G_\theta(\mathbf{k};t',t')=1,
\label{Gtheta eq. initial}
\end{equation}
which are consistent with the rigorous relations $\Theta(\mathbf{k};t,t')=\Theta(\mathbf{k};t',t')$ and $G_\theta(\mathbf{k};t,t')=1$ for the limit case $\kappa\to 0$. 
\fi

%In a limit case of $T^2 \to \infty$ ($\xi\to\infty$), Eqs. (\ref{composite H})-(\ref{composite G}) read $H\to\Theta$ and $G_H\to G_\theta$ whose governing equations (\ref{ETC LRA'})-(\ref{G LRA'}) tend to those of scalar-based LRA \cite{Kaneda86}. 

%Equations (\ref{composite H}) and (\ref{composite G}) tell us $H(\mathbf{k};t,t')\to\Theta(\mathbf{k};t,t')$ and $G_H(\mathbf{k};t,t')\to G_\theta(\mathbf{k};t,t')$ for a limiting case $n\to\pm \infty$ ($\alpha(n)\to 0$), then a closed set of Eqs. (\ref{ETC LRA'})-(\ref{G LRA'}) exactly tend to what is obtained by scalar-based LRA of Ref. \cite{}. On the contrary, $n=1/3$ (traceless Hessian) completely omits $\Theta(\mathbf{k};t,t')$ and $G_\theta(\mathbf{k};t,t')$ from Eqs. (\ref{composite H}) and (\ref{composite G}) for $\alpha(n)=3/2$, giving a closure model purely on the basis of the traceless Hessian.

\section{Application to inertial convective range}\label{INERTIAL CONVECTIVE}
\subsection{HBLRA for isotropic turbulence}
For homogeneous isotropic cases, all the statistical functions take isotropic forms;
\begin{subequations}
\begin{align}
&H(\mathbf{k};t,t')=H(k;t,t'),\\
&G_H(\mathbf{k};t,t')=G_H(k;t,t'),\\
&Q_{ij}(\mathbf{k};t,t')=\frac{1}{2}P_{ij}(\mathbf{k})Q(k;t,t'),
\end{align}
\end{subequations}
where $k=\lVert\mathbf{k}\rVert$. The resultant equations for $H(k;t,t')$ and $G_H(k;t,t')$ are
\begin{equation}
\begin{split}
&\left(\partial_t+2\kappa k^2\right)H(k;t,t)\\
&=2\pi\iint_\triangle dp\,dq\, kpq(1-z^2)\int^t_{t_0}ds
Q(p;t,s)\left\{H(q;s,t)G_H(k;t,s)-H(k;s,t)G_H(q;t,s)\right\},
\end{split}
\label{HI ETC eq.}
\end{equation}

\begin{equation}
\begin{split}
&\!\!\!\!\!\!\!\left(\partial_t+\kappa k^2\right) H(k;t,t')\\
=&-\frac{\pi}{T^2}\iint_\triangle dp\,dq\, kpq(1-y^2)(1-z^2)\int^t_{t'}ds\,Q(p;t,s) H(k;t,t')\\
&-\frac{\pi}{T^2}\iint_\triangle dp\,dq\, kpq(1-y^2)(1-z^2)\int^t_{t_0}ds\,Q(p;t,s) G_H(q;t,s)H(k;s,t')\\
&+\frac{\pi}{T^2}\iint_\triangle dp\,dq\, kpq(1-y^2)(1-z^2)\int^{t'}_{t_0}ds\,Q(p;t,s) H(q;t,s)G_H(k;t',s),
\end{split}
\label{HI 2TC eq.}
\end{equation}

\begin{equation}
\begin{split}
&\left(\partial_t+\kappa k^2\right)G_H(k;t,t')\\
&=-\frac{\pi}{T^2}\iint_\triangle dp\, dq\, kpq(1-y^2)(1-z^2)\\
&\ \ \ \ \ \ \ \ \ \ \ \ \ \ \ \ 
\times \int^t_{t'}ds\, Q(p;t,s)
\left\{G_H(k;t,t')+G_H(q;t,s)G_H(k;s,t')\right\}\ \ (t\geq t'),
\end{split}
\label{HI GH eq.}
\end{equation}

\begin{equation}
G_H(k;t',t')=1,
\label{HI GH initial}
\end{equation}
where the geometrical factors $y\equiv (q^2+k^2-p^2)/(2kq)$, $z\equiv (q^2+k^2-p^2)/(2kq)$, and $\triangle\equiv\{(p,q)|\ |k-p|\leq q \leq k+p \}$ reflect the triad interaction between three wavenumber modes; second-order nonlinearity allows an interaction between three modes when $k$, $p$, and $q$ can form the legs of a triangle. Then three factors $x(\equiv (p^2+q^2-k^2)/(2pq))$, $y$, and $z$ are introduced as cosines of three interior angles opposite the legs $k$, $p$, and $q$, respectively \cite{Kraichnan59,Kaneda81}. The integration domain $\triangle$ arises from an existence condition for such triangle.

Also note that a common geometrical factor $kpq(1-y^2)(1-z^2)$ appears in the wavenumber integration of Eqs. (\ref{HI 2TC eq.}) and (\ref{HI GH eq.}). This is indeed equivalent to the one in the time-scale integral of the velocity-closure model of LRA (see Eqs. (2.50) and (2.52) of Ref. \cite{Kaneda81}). This geometrical factor sufficiently reduces both large and small scale contributions, which is the very key to reproducing scale-local interaction consistent with the Kolmogorov-Obukhov-Corrsin theory.

\subsection{Inertial-convective range}
\begin{figure}
\centering
\includegraphics[width=10cm]{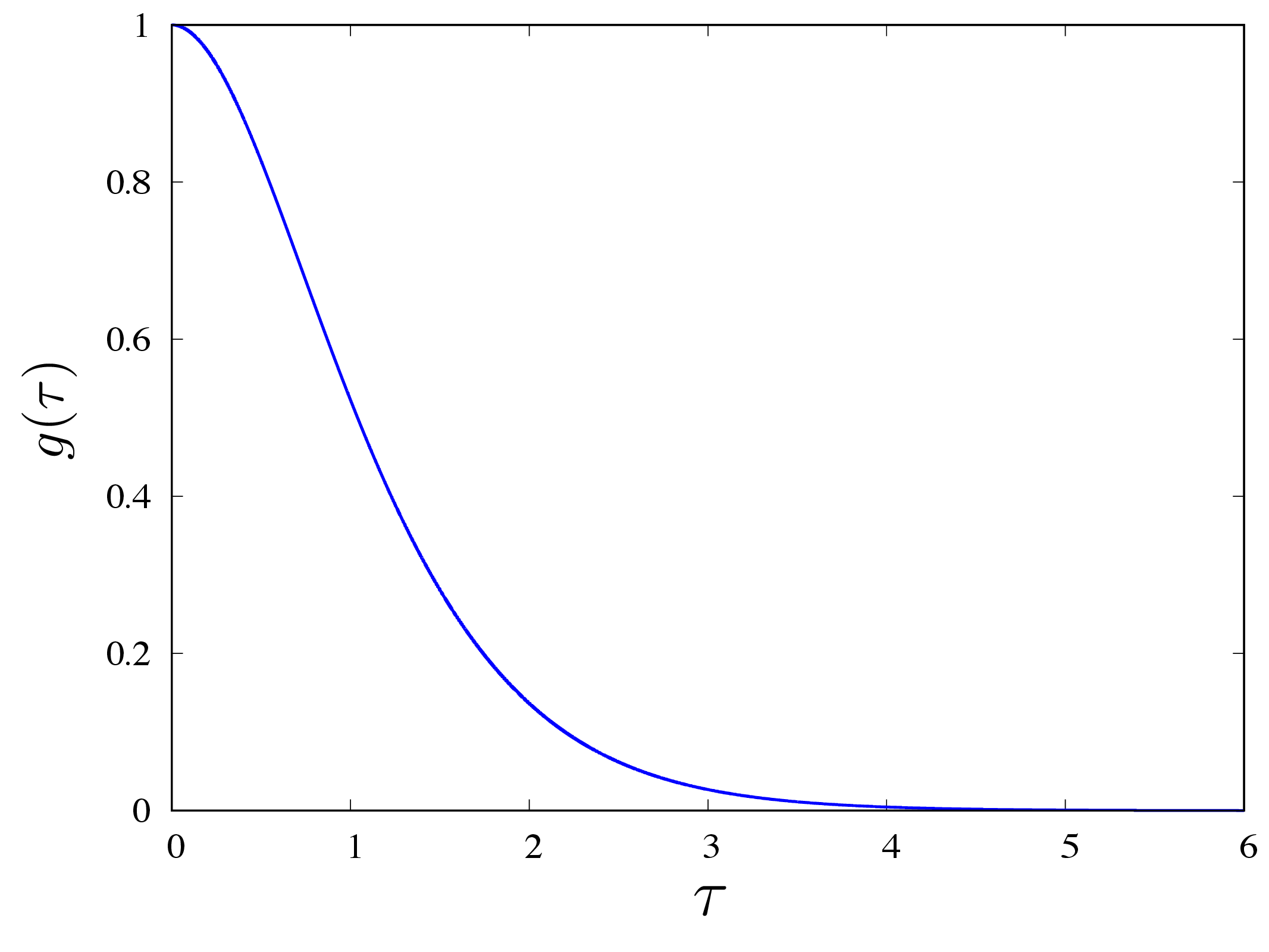}
\caption{Configuration of the dimensionless function $g(\tau)$ obtained from LRA equations of Ref. \cite{Kaneda86}. $g(\tau)$ is monotonically decaying as the dimensionless time $\tau$ passes.}
\label{gtau}
\end{figure}

\begin{figure}
\centering
\includegraphics[width=10cm]{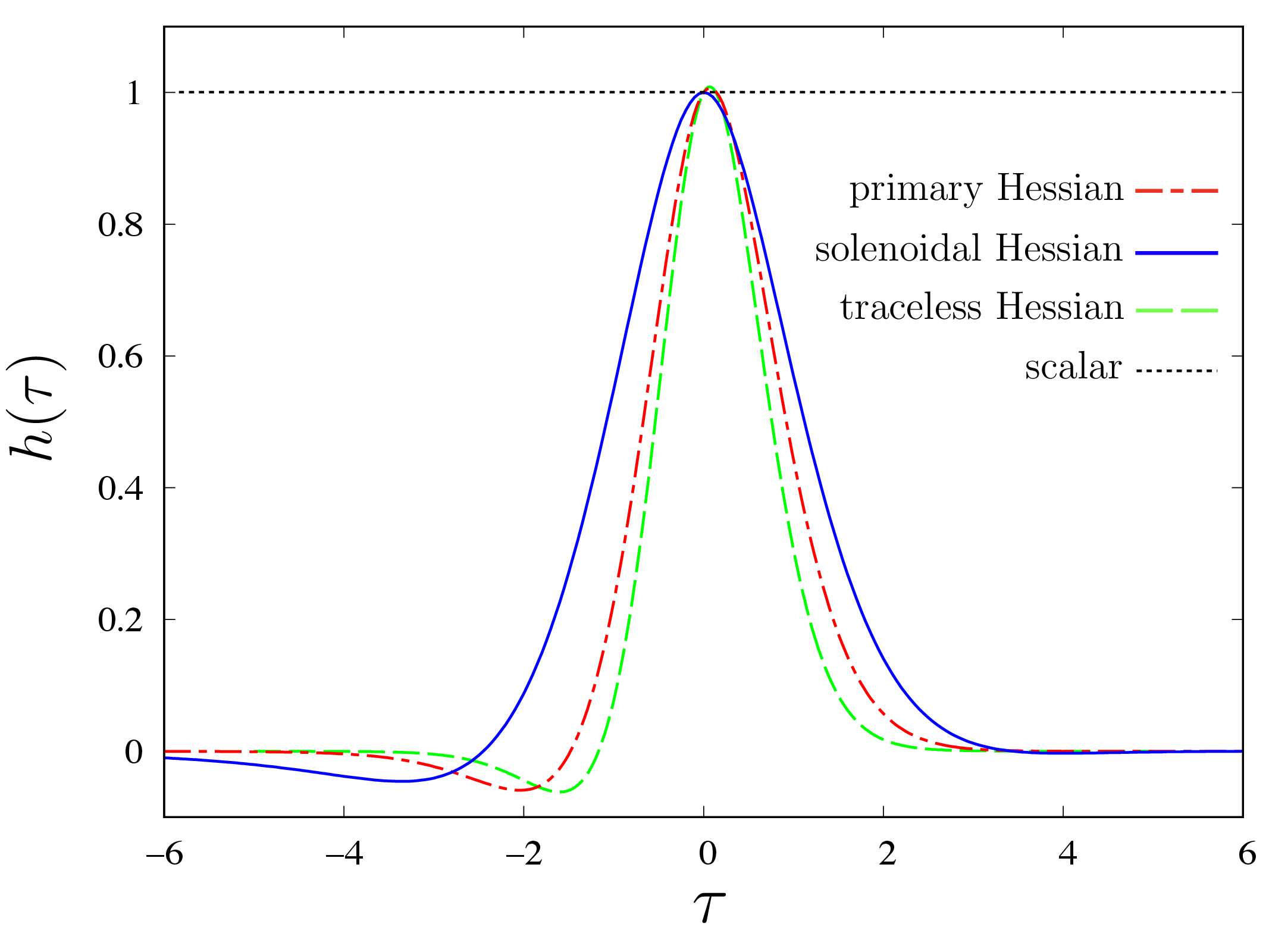}
\caption{Configuration of the universal function $h(\tau)$ which rapidly decays as $|\tau|$ increases. Unlike the velocity auto-correlation, $h(\tau)$ is asymmetric in time reversing.}
\label{h config}
\end{figure}

\begin{figure}
\centering
\includegraphics[width=10cm]{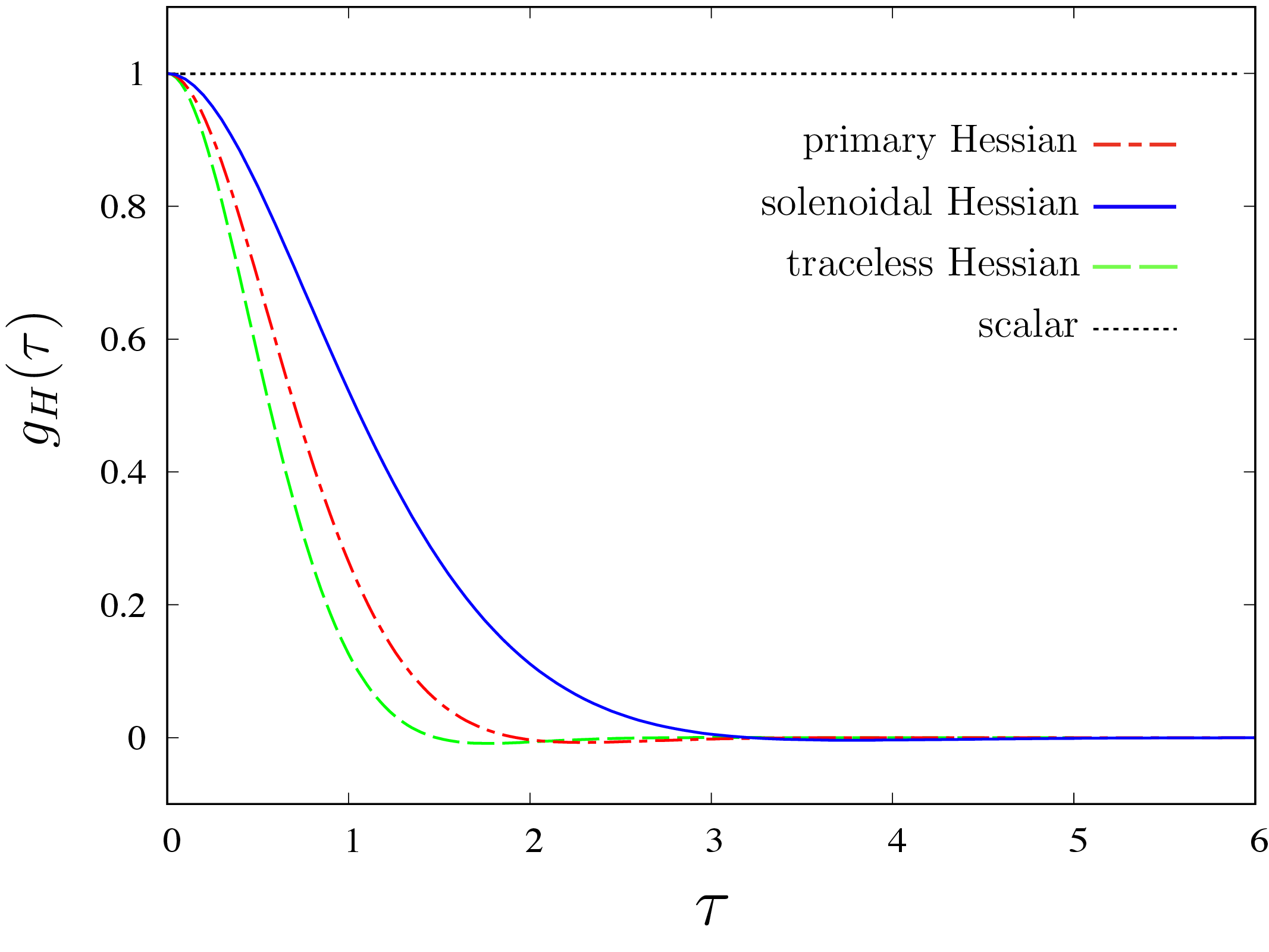}
\caption{Configuration of the universal function $g_H(\tau)$ which rapidly decays within finite time scale. Unlike the scalar field preserving long-time memory, Hessian field loses its memory in the inertial-range time scale.}
\label{Ggh config}
\end{figure}

Let us apply Eqs. (\ref{HI 2TC eq.})-(\ref{HI GH initial}) to the inertial-convective range. We first assume both velocity and scalar fields are sustained by some source at sufficiently large scale while viscosity and diffusivity act at very small scale, so that a broadband inertial-convective range is realized under quasi-stationary state. Scale-locality of velocity-scalar coupling brings about constant spectral flux of the scalar variance, yielding universal scaling law of scalar-variance spectrum in a parallel manner to the LRA analysis of Ref. \cite{Kaneda86} where the inertial-range solution of $Q$ had been obtained:
\begin{equation}
Q(k;t,t')=\frac{K_o}{2\pi}\bepsilon^{2/3}k^{-11/3}g(\bepsilon^{1/3}k^{2/3}(t-t')),
\label{K86 solution}
\end{equation}
where $K_o\approx 1.72$ is the Kolmogorov constant theoretically obtained from LRA analysis, $\bepsilon=\nu\langle (\partial_i u_j)(\partial_i u_j)\rangle$ is the mean dissipation rate of energy, and $g$ is a dimensionless function given by Fig. \ref{gtau} ($g(0)=1$). Now we shall solve the scalar statistics $H$ and $G_H$ using Eqs. (\ref{HI ETC eq.})-(\ref{HI GH eq.}) of the present HBLRA framework. A simple dimensional analysis applied to Eq. (\ref{HI GH eq.}) tells us that $G_H(k;t,t')$ has the time scale of the inertial range, i.e. $\bepsilon^{-1/3}k^{-2/3}$ while the same analysis holds for $H(k;t,t')$ in Eq. (\ref{HI 2TC eq.}). Thus the following may be allowed as the demanded solution of Eqs. (\ref{HI ETC eq.})-(\ref{HI GH initial}) in the inertial-convective range:
\begin{subequations}
\begin{align}
H(k;t,t')&=\frac{K_\theta}{4\pi}
\bepsilon^{-1/3}\bchi k^{-11/3}h(\bepsilon^{1/3}k^{2/3}(t-t')),\\ 
G_H(k;t,t')&=g_H(\bepsilon^{1/3}k^{2/3}(t-t')),
\end{align}
\label{scale similar}
\end{subequations}
where $\bchi=\kappa\langle (\partial_i \theta)(\partial_i \theta)\rangle$, $h(\tau)$ and $g_H(\tau)$ are dimensionless functions ($h(0)=g_H(0)=1$). In what follows, we consider quasi stationary state truly independent from the initial fields at $t_0$, so the initial time $t_0$ in Eqs. (\ref{HI ETC eq.})-(\ref{HI GH eq.}) is to be taken as an infinite past time, i.e. $t_0\to -\infty$. Substituting the scale-similar solutions (\ref{scale similar}) into Eqs. (\ref{HI 2TC eq.})-(\ref{HI GH eq.}) yields
\begin{equation}
\begin{split}
\frac{d}{d\tau}h(\tau)
=&-\frac{K_o}{2T^2}\iint_{\hat{\triangle}}d\hat{p}\,d\hat{q}\,
\hat{p}^{-8/3}\hat{q}(1-y^2)(1-z^2)\int^\tau_0d\sigma\,g(\hat{p}^{2/3}\sigma)h(\tau)\\
&-\frac{K_o}{2T^2}\iint_{\hat{\triangle}}d\hat{p}\,d\hat{q}\,
\hat{p}^{-8/3}\hat{q}(1-y^2)(1-z^2)\int^\infty_0d\sigma\,g(\hat{p}^{2/3}\sigma)
g_H(\hat{q}^{2/3}\sigma)h(\tau-\sigma)\\
&+\frac{K_o}{2T^2}\iint_{\hat{\triangle}}d\hat{p}\,d\hat{q}\,
\hat{p}^{-8/3}\hat{q}^{-8/3}(1-y^2)(1-z^2)\int^\infty_0d\sigma\,g(\hat{p}^{2/3}(\tau+\sigma))
h(\hat{q}^{2/3}(\tau+\sigma))g_H(\sigma),
\end{split}
\label{inertial 2TC}
\end{equation}

\begin{equation}
\begin{split}
\frac{d}{d\tau}g_H(\tau)=-\frac{K_o}{2T^2}\iint_{\hat{\triangle}}&d\hat{p}\,d\hat{q}\,
\hat{p}^{-8/3}\hat{q}(1-y^2)(1-z^2)\\
&\times\int^\tau_0 d\sigma\,g(\hat{p}^{2/3}\sigma)
\left\{g_H(\tau)+g_H(\hat{q}^{2/3}\sigma)g_H(\tau-\sigma)\right\}\ \ (\tau\geq 0),
\end{split}
\label{inertial GH}
\end{equation}
where $\hat{p},\hat{q}\equiv p/k,q/k$. Here note that $K_\theta$ disappears from the above equations, so $h(\tau)$ and $g_H(\tau)$ can be solved irrespective of $K_\theta$. With the help of $K_o\approx 1.72$ and $g(\tau)$ obtained by LRA analysis \cite{Kaneda86} and $h(0)=g_H(0)=1$, the dimensionless Eqs. (\ref{inertial 2TC}) and (\ref{inertial GH}) can be numerically solved in terms of $h(\tau)$ and $g_H(\tau)$ whose configurations are given respectively by Figs. \ref{h config} and \ref{Ggh config}. These dimensionless functions characterizes time scales of the representative variables in the unit of the inertial-range time scale $\bepsilon^{-1/3}k^{-2/3}$: 

\begin{equation}
I_h=\frac{1}{2}\int^\infty_{-\infty} h(\tau) d\tau \approx 
\begin{dcases}
&0.816\ (\textrm{primary Hessian}),\\
&1.12\ (\textrm{solenoidal Hessian}),\\
&0.664\ (\textrm{traceless Hessian}),\\
&\infty\ (\textrm{scalar}).
\end{dcases}
\label{Ih comparison}
\end{equation}

\begin{equation}
I_g=\int^\infty_0 g_H(\tau) d\tau \approx
\begin{dcases}
&0.748\ (\textrm{primary Hessian}),\\
&1.13\ (\textrm{solenoidal Hessian}),\\
&0.595\ (\textrm{traceless Hessian}),\\
&\infty\ (\textrm{scalar}),
\end{dcases}
\label{Ig comparison}
\end{equation}
which are to be compared with $\int^\infty_0 g(\tau) d\tau = 1.19$ \cite{Kaneda86}; i.e. Hessian field generally gives shorter correlation time scales than that of velocity field. The traceless Hessian loses its memory most rapidly. This is because the traceless Hessian is completely free from the long-time memory of the scalar $\theta$. The other two may be rewritten on the basis of the traceless one combined with the scalar $\theta$ as their trace parts;  
\begin{equation}
\mathscr{H}^\mathrm{(p)}_{ij}=\mathscr{H}^\mathrm{(t)}_{ij}+\frac{1}{3}\theta\delta_{ij},
\end{equation}
\begin{equation}
\mathscr{H}^\mathrm{(s)}_{ij}=\mathscr{H}^\mathrm{(t)}_{ij}-\frac{2}{3}\theta\delta_{ij},
\end{equation}
where $(t)$, $(p)$, and $(s)$ stands for ``traceless", ``primary", and ``solenoidal" (see Sec. \ref{HESSIAN FIELD}).  The solenoidal Hessian has longer timescale than the primary one for its larger weight on $\theta$.

Now the dimensionless functions $h(\tau)$ and $g_H(\tau)$ had been solved from Eqs. (\ref{HI 2TC eq.})-(\ref{HI GH initial}), then we shall go for the Obukhov-Corrsin constant $K_\theta$ which is to be obtained from the rest Eq. (\ref{HI ETC eq.}). Note that, however, both sides of Eq. (\ref{HI ETC eq.}) trivially vanishes in the inertial-convective range, and we should transform this equation into the spectral flux so that a non-trivial equation for $K_\theta$ may be obtained. Multiplying Eq. (\ref{HI ETC eq.}) by $4\pi k^2$ yields the dynamical equation for the scalar-variance spectrum $E_\theta(k)$ $(\equiv 4\pi k^2 H(k;t,t))$:
\begin{equation}
(\partial_t+2\kappa k^2)E_\theta(k,t)=T_\theta(k,t), 
\label{Etheta 0}
\end{equation} 
where the scalar-transport function $T_\theta(k,t)$ reads
\begin{equation}
T_\theta(k,t)=\frac{1}{2}\iint_\triangle dp\,dq\, S_\theta(k;p,q|t),
\label{Ttheta0}
\end{equation}
\begin{equation}
\begin{split}
S_\theta(k;p,q|t)=8\pi^2 k^3pq\int^t_{-\infty}ds\Bigg[
 &(1-z^2) Q(p;t,s)\left\{H(q;s,t)G_H(k;t,s)-H(k;s,t)G_H(q;t,s)\right\}\\
+&(1-y^2) Q(q;t,s)\left\{H(p;s,t)G_H(k;t,s)-H(k;s,t)G_H(p;t,s)\right\}\Bigg].
\end{split}
\label{triad interaction}
\end{equation}
Now the triad interaction between three modes $k$, $p$, and $q$ are mediated by $S_\theta(k;p,q|t)$ which is symmetric under the exchange between $p$ and $q$; i.e. $S_\theta(k;p,q|t)=S_\theta(k;q,p|t)$. Then the integral domain $\triangle$ can be reduced to its half:
\begin{equation}
T_\theta(k,t)=\int^\infty_0 dp\int^{p+k}_{\mathrm{max}(p,k-p)} dq\, S_\theta(k;p,q|t).
\label{Ttheta1}
\end{equation}
Due to the sine theorem $(1-x^2)/k^2=(1-y^2)/p^2=(1-z^2)/q^2$, detailed conservation holds:
\begin{equation}
S_\theta(k;p,q|t)+S_\theta(p;q,k|t)+S_\theta(q;k,p|t)=0,
\label{d balance}
\end{equation}
which guarantees the conservation of the scalar variance $\langle\theta^2\rangle$ in the limit $\kappa\to\infty$. The spectral flux is defined by 
\begin{equation}
\Pi_\theta(k,t)=\int^\infty_k dk' 
\int^\infty_0 dp'\int^{p'+k'}_{\mathrm{max}(p',k'-p')} dq'\, S_\theta(k';p',q'|t).
\label{Pi0}
\end{equation}
In the inertial-convective range, this scalar flux may balance with $\bchi$. For further progress we should calculate $\Pi_\theta(k,t)$ of Eq. (\ref{Pi0}) with Eqs. (\ref{K86 solution})-(\ref{scale similar}) substituted, which, however, is an integral over an infinitely large domain in the wavenumber space. Following a similar procedure provided by Ref. \cite{Kraichnan66} on the velocity statistics, we rewrite Eq. (\ref{Pi0}) as an integral over a finite domain (see Appendix \ref{SCALE-SIMILAR PI} for its derivation):
\if0
Due to the detailed conservation (\ref{d balance}), the integration (\ref{Pi0}) over the range $\{(k',p',q')|k',p',q'>k\}$ vanishes. Then Eq. (\ref{Pi0}) is reduced to
\begin{equation}
\Pi_\theta(k,t)=\int^\infty_k dk' 
\int^k_0 dp'\int^{p'+k'}_{\mathrm{max}(p',k'-p')} dq'\, S_\theta(k';p',q'|t).
\label{Pi}
\end{equation}
In the inertial-convective range, a scale-similar transformation $S_\theta(k';p',q'|t)=a^3 S_\theta(ak';ap',aq'|t)$ holds. Here we choose $a=k/k'$, $b=ap'$, and $q=aq'$, so that Eq. (\ref{Pi}) becomes
\begin{equation}
\Pi_\theta(k,t)=\int^\infty_k dk' \frac{k}{k'}
\int^{k^2/k'}_0 db\int^{b+k}_{\mathrm{max}(b,k-b)} dq\, S_\theta(k;b,q|t),
\label{Pi1}
\end{equation}
By setting $p=k^2/k'$ Eq. (\ref{Pi1}) turns into
\begin{equation}
\Pi_\theta(k,t)=\int^k_0 dp\ \frac{k}{p}
\int^p_0 db\int^{b+k}_{\mathrm{max}(b,k-b)} dq\, S_\theta(k;b,q|t).
\label{Pi1}
\end{equation}
Partial integration by $p$ yields
\fi
\begin{equation}
\Pi_\theta(k,t)
=k\int^k_0 dp\, \mathrm{ln}\left(\frac{k}{p}\right)
\int^{p+k}_{\mathrm{max}(p,k-p)} dq\, S_\theta(k;p,q|t).
\label{Pi3}
\end{equation}
\begin{figure}
\centering
\includegraphics[width=10cm]{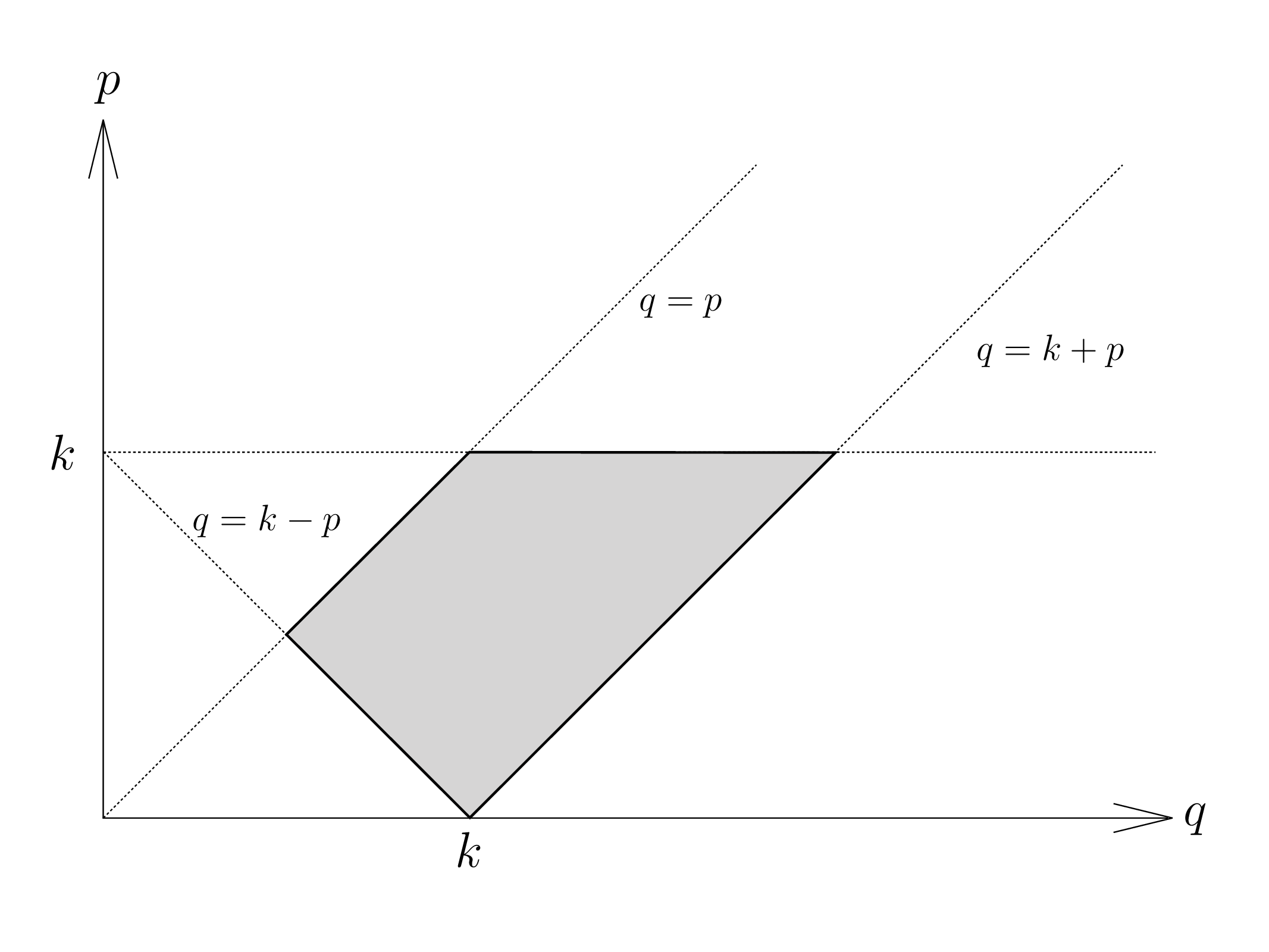}
\caption{Integration domain of Eq. (\ref{Pi3}) in the $p$-$q$ space.}
\label{intgdomain}
\end{figure}
Now the integration domain is limited in the gray area in Fig. \ref{intgdomain}. Then a scale-similar transformation $S_\theta(k;p,q|t)=k^{-3}S_\theta(1;\hat{p},\hat{q}|t)$ ($\hat{p}\equiv p/k,\hat{q}\equiv q/k$) further simplifies the above (see appendix \ref{SCALE-SIMILAR PI}), leading to
\begin{equation}
\Pi_\theta(k,t)=\int^1_0 d\hat{p}\, \mathrm{ln}\left(\frac{1}{\hat{p}}\right)
\int^{\hat{p}+1}_{\mathrm{max}(\hat{p},1-\hat{p})} d\hat{q}\, S_\theta(1;\hat{p},\hat{q}|t).
\label{constant Pi}
\end{equation}
Now the right side is apparently independent from $k$, suggesting that the spectral flux $\Pi_\theta(k,t)$ takes a constant value in the inertial-convective range. Substituting Eq. (\ref{triad interaction}) and scale-similar forms (\ref{K86 solution})-(\ref{scale similar}) into Eq. (\ref{constant Pi}), the scalar-flux relation $\Pi_\theta = \bchi$ reads
\begin{equation}
\begin{split}
\Pi_\theta(k,t)=K_oK_\theta\bchi\int^1_0 d\hat{p}\ &\mathrm{ln}\left(\frac{1}{\hat{p}}\right)
\int^{\hat{p}+1}_{\mathrm{max}(\hat{p},1-\hat{p})} d\hat{q}\int^\infty_0d\tau\\
\times\Bigg[
 &(1-z^2)\hat{p}^{-8/3}\hat{q}\ g(\hat{p}^{2/3}\tau)
\left\{\hat{q}^{-11/3}h(-\hat{q}^{2/3}\tau)g_H(\tau)-h(-\tau)g_H(\hat{q}^{2/3}\tau)\right\}\\
+&(1-y^2)\hat{q}^{-8/3}\hat{p}\ g(\hat{q}^{2/3}\tau)
\left\{\hat{p}^{-11/3}h(-\hat{p}^{2/3}\tau)g_H(\tau)-h(-\tau)g_H(\hat{p}^{2/3}\tau)\right\}
\Bigg]\\
=\bchi,\ \ \ \ \ \ \ \ \ \ \ \ \ \ \ \ \ &
\end{split}
\label{OC equation}
\end{equation}
which is now solvable in terms of $K_\theta$. Substituting $K_o\approx 1.72$ \cite{Kaneda86} into the above yields 
\begin{equation}
K_\theta\approx
\begin{dcases}
&1.03\ (\textrm{primary Hessian}),\\
&0.754\ (\textrm{solenoidal Hessian}),\\
&1.23\ (\textrm{traceless Hessian}),\\
&0.337\ (\textrm{scalar}).
\end{dcases}
\label{OC constant}
\end{equation}
where the result for scalar is identical to the one obtained by Ref. \cite{Kaneda86} ($K_\theta \approx 0.34$). The variation of the constant value can be understood from that of timescale of the memory functions $h$ and $g_h$ in Eq. (\ref{OC equation}); shorter memory timescale regulates the magnitude of the scalar flux $\Pi_\theta$ via time integration of $h$ and $g_h$, then the resultant $K_\theta$ become larger. The memory timescales are given, in ascending order, by traceless, primary, and solenoidal Hessians which give $K_\theta$ in descending order ( see Eqs. (\ref{Ig comparison}) and (\ref{Ih comparison}) ). In comparison with experimental and numerical assessments ($K_\theta$ approximately ranges from 0.6 to 0.9 in pioneering works \cite{Sreeni96, MW98, WCB99, YXS02, WG04, GW15}), Hessian-based analyses predict reasonable values of $K_\theta$. In case of scalar-based LRA, infinitely-long memory effect then underestimates $K_\theta$.

\subsection{Locality of the interscale interaction}
Besides a quantitative improvements on the Obukhov-Corrsin constant $K_\theta$, a qualitative difference should be noted in the physical interpretation of the scalar flux $\Pi_\theta$. The right side of Eq. (\ref{OC equation}) tells us detailed contributions from velocity and scalar of various modes to $\Pi_\theta$. Let us focus on the low-wavenumber contribution at $\hat{p}(=p/k)\ll 1$ which corresponds to relatively low-wavenumber mode $p(\ll k)$ still within the inertial-convective range. In this scale range, the first among two integrands becomes dominant, where $g(\hat{p}^{2/3}\tau)$ represents a slow decorrelation of larger eddy (still within the inertial range) in comparison with Hessian statistics ( $h(-\hat{q}^{2/3}\tau)$, $g_H(\tau)$, $h(-\tau)$, and $g_H(\hat{q}^{2/3}\tau)$ ) of finite-time decorrelation. Then the following approximately holds:
\begin{equation}
\begin{split}
&\int_0^\infty d\tau g(\hat{p}^{2/3}\tau)
\left\{\hat{q}^{-11/3}h(-\hat{q}^{2/3}\tau)g_H(\tau)-h(-\tau)g_H(\hat{q}^{2/3}\tau)\right\}\\
&\approx
\int_0^\infty d\tau
\left\{\hat{q}^{-11/3}h(-\hat{q}^{2/3}\tau)g_H(\tau)-h(-\tau)g_H(\hat{q}^{2/3}\tau)\right\},
\end{split}
\label{infrared2pi}
\end{equation} 
implying that the memory of the large-scale fluid's motion represented by $g(\hat{p}^{2/3}\tau)$ may be screened out by the scalar decorrelation at mode $k$ (note that $p\ll k\approx q$). In case of scalar-based LRA, by contrast, this term becomes
\begin{equation}
\left(\hat{q}^{-11/3}-1\right)\int_0^\infty d\tau g(\hat{p}^{2/3}\tau) 
\end{equation}
where long-time memory of the velocity field from mode $p (\ll k)$ may be cast into the scalar flux at mode $k$, precisely due to the absence of the memory-fading effect of the scalar statistics at mode $k$. While being consistent with Kolmogorov-Obukhov-Corrsin dimensional analysis, scalar-based LRA may predicts more broadband interaction than HBLRA via interference from turbulence eddy at larger scale. 
%Such a broadband interaction in the inertial range may well overestimate the scalar cascade and, eventually, underestimate the Obukhov-Corrsin constant $K_\theta$. 

\begin{figure}
\centering
\includegraphics[width=10cm]{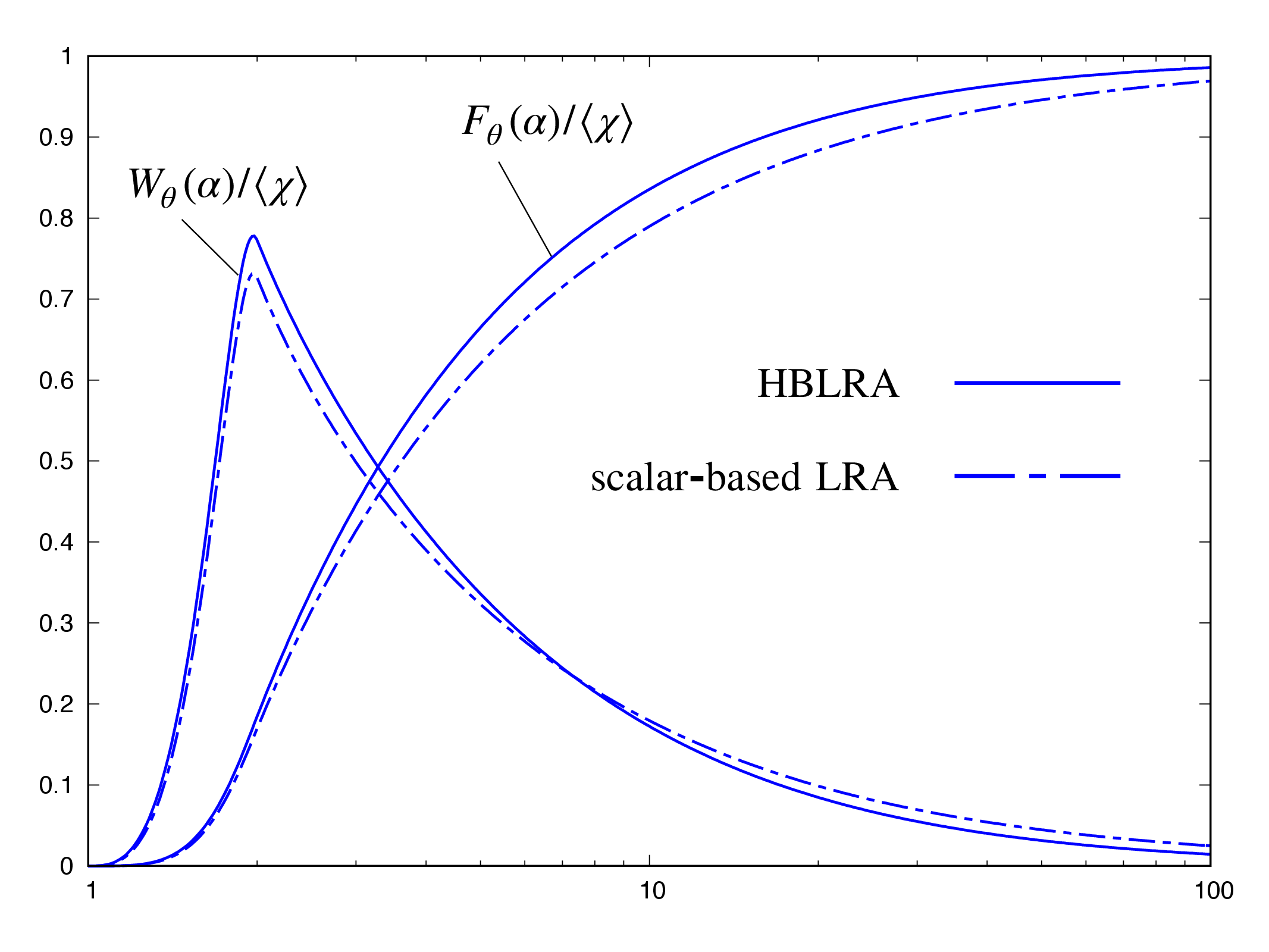}
\caption{$W_\theta(\alpha)$ peaks at $\alpha\approx 2$ while $F_\theta(\alpha)$ monotonically increases tending to unity as $\alpha\to \infty$. Their asymptotic behaviors $W_\theta/\langle\chi\rangle\sim 1-F_\theta/\langle\chi\rangle \sim  \alpha^{-4/3}\mathrm{ln}\,\alpha$ (HBLRA) and $W_\theta/\langle\chi\rangle\sim 1-F_\theta/\langle\chi\rangle \sim  \alpha^{-2/3}\mathrm{ln}\,\alpha$ (scalar-based LRA) are obtained.
}
\label{locality}
\end{figure}

\begin{figure}
\centering
\includegraphics[width=17cm]{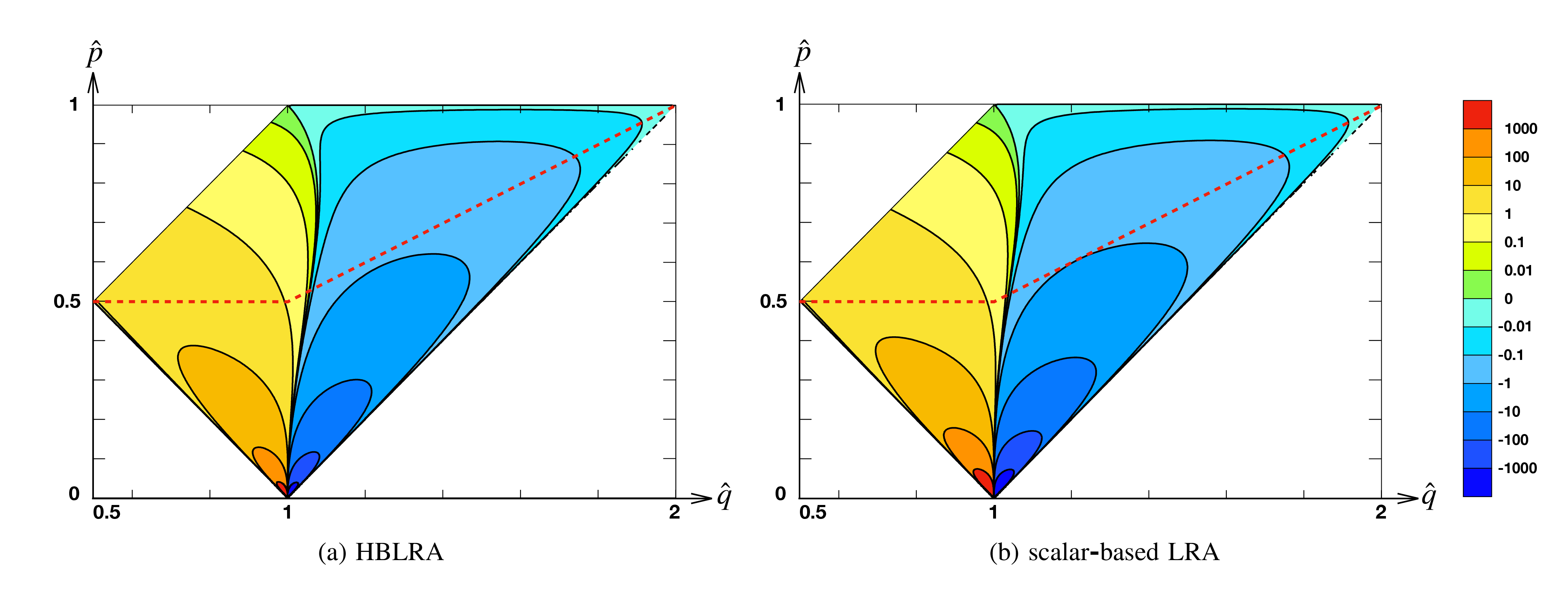}
\caption{Distributions of integrand in Eq. (\ref{constant Pi}) for (a) HBLRA and (b) scalar-based LRA. More divergent distribution is observed around $(\hat{q},\hat{p})=(1,0)$ in (b), where $Q(p;t,s)$ in Eq. (\ref{triad interaction}) diverges as $\hat{p}^{-11/3}$. Red dotted lines show $\alpha=\max(1,\hat{p},\hat{q})/\min(1,\hat{p},\hat{q})=2$. }
\label{S_distribution}
\end{figure}

To see more deeper insights, here we quantify the locality (or the non-locality) of the interscale interaction in the inertial range following Refs. \cite{Kraichnan66,Kaneda86}. We may introduce the scale-locality parameter $\alpha$ of each triad $(k,p,q)$ by 
\begin{equation}
\alpha\equiv \frac{\mathrm{max}(k,p,q)}{\mathrm{min}(k,p,q)} 
= \frac{\mathrm{max}(1,\hat{p},\hat{q})}{\mathrm{min}(1,\hat{p},\hat{q})}.
\end{equation}
Now Eq. (\ref{constant Pi}) may be rewritten as an integration over $1\leq\alpha\leq\infty$: 
\begin{equation}
\Pi_\theta = \int_1^\infty W_\theta(\alpha) \frac{d\alpha}{\alpha},
\end{equation}
where
\begin{equation}
\begin{split}
W_\theta(\alpha) =& \frac{\mathrm{ln}\alpha}{\alpha}\int^{\alpha^*}_1 
S_\theta\left(1;\frac{1}{\alpha},\frac{1}{\beta}\right)\frac{d\beta}{\beta^2}
+\alpha\int^\alpha_{\alpha^{**}} S_\theta\left(1;\frac{1}{\beta},\frac{\alpha}{\beta}\right)
\frac{d\beta \mathrm{ln\beta}}{\beta^3},
\end{split}
\end{equation}

\begin{equation}
\alpha^* = \mathrm{min}\left(\alpha,\frac{\alpha}{\alpha-1}\right),\ \ 
\alpha^{**} = \mathrm{max}\left(1,\alpha-1\right).
\end{equation}
Now $W_\theta(\alpha)$ measures the weight of triad interaction where the ratio of maximum to minimum wavenumber is $\alpha$. Figure \ref{locality} shows $W_\theta(\alpha)$ and $F_\theta(\alpha)(\equiv \int_1^{\alpha}W_\theta(\alpha')d\alpha')$ normalized by $\bchi$. Here we employs the case of solenoidal Hessian representing HBLRA since very similar distributions are obtained for the rest primary and traceless. As discussed in Refs. \citep{Kraichnan66, Kaneda86} for velocity field, triad modes of $\alpha\approx 2$ may be most prominent among all interactions; namely, an event where some scalar distribution is broken down to its half size may be frequently observed, which, however, does not necessarily mean that this event dominates the total scalar transfer. HBLRA explicates via $F_\theta(\alpha)$ that interactions of $\alpha\leq 2$ occupy only 19 \%  of the total scalar transfer (21 \% in case of the velocity field) and the rest 81 \% by the tail of $\alpha\geq 2$. Whereas being scale-local enough to yield Kolmogorov-Obukhov-Corrsin scaling, this somewhat broadband interaction requires certainly wide inertial-convective range before reaching at the true universality (see similar discussion by Ref. \citep{Kraichnan66} for velocity field). In addition, $W_\theta$ peaks at $0.78$ slightly lower than $0.87$ of the velocity; the scalar field may represent somewhat weaker scale-locality than the velocity field. HBLRA predicts $W_\theta(\alpha) \sim \alpha^{-4/3} \mathrm{ln}\,\alpha$ for $\alpha\to \infty$ which coincides with that of velocity field \citep{Kraichnan66}. Now $1 - F_\theta(\alpha)\sim \alpha^{-4/3} \mathrm{ln}\,\alpha$ ($\alpha\to\infty$) and $F_\theta(\alpha)$ reaches $0.99\,\bchi$ at $\alpha \approx 140$; i.e., 99 \% of the scalar flux at a mode $k$ comes from a spectral band $k/140 \sim 140\times k$. Providing $k_s$ and $k_\ell$ represent the smallest and largest scales of the inertial-convective range, Eq. (\ref{OC equation}) implies that the scale gap of $k_s/k_\ell\gtrsim 140^2 \approx 2 \times 10^4$ may be necessary to obtain the Obukhov-Corrsin constant within the error of a few percent (an identical analysis on the Kolmogorov constant results in $k_s/k_\ell\gtrsim 1\times 10^4$). On the other hand, the scalar-based LRA represents more broadband interaction due to longer tail of $W_\theta(\alpha)\sim\alpha^{-2/3} \mathrm{ln}\,\alpha$ ($\alpha\to\infty$) caused by large-scale mode of the velocity field. Then $1-F_\theta \sim \alpha^{-2/3} \mathrm{ln}\,\alpha$ and $F_\theta(\alpha)$ reaches $0.99\,\bchi$ when $\alpha \approx 400$, which corresponds to the scale gap of $k_s/k_\ell\gtrsim 1.6 \times 10^5$. The difference of tails exactly caused by interference from large-scale eddy, which may be elucidated by distribution of the integrand of Eq. (\ref{constant Pi}) given by Fig. \ref{S_distribution}. In both (a) HBLRA and (b) scalar-based LRA, most of the positive (negative) contribution comes from $\hat{q}\lesssim 1$ ($\hat{q}\gtrsim 1$), which implies the positive (negative) scalar transfer from smaller (larger) $q$ mode to $k$ mode. Triad interactions of $\alpha\geq 2$ corresponds to the region lower than the red dotted line where the scale-non-local interaction ($p\ll q \approx k$) may be observed in the vicinity of $(\hat{q},\hat{p})=(1,0)$. In case of (a) HBLRA, $S_\theta(1;\hat{p},\hat{q})$ behaves as $\hat{p}^{-8/3}$ for $\hat{p}\to 0$, while (b) scalar-based LRA shows more divergent trend; $S_\theta(1;\hat{p},\hat{q})\sim\hat{p}^{-10/3}$. These asymptotic behaviors originate from divergence of $Q(p;t,s)$ in Eq. (\ref{triad interaction}) for $p\to 0$, not from the scalar properties; HBLRA reduces convection effect of large-scale eddy, precisely due to rapid memory-fading of scalar field. \\\\

%Note that such a trend may be also seen in other  branch of Lagrangian closures (LHDIA, SBALHDIA, etc.).

\begin{figure}
\centering
\includegraphics[width=10cm]{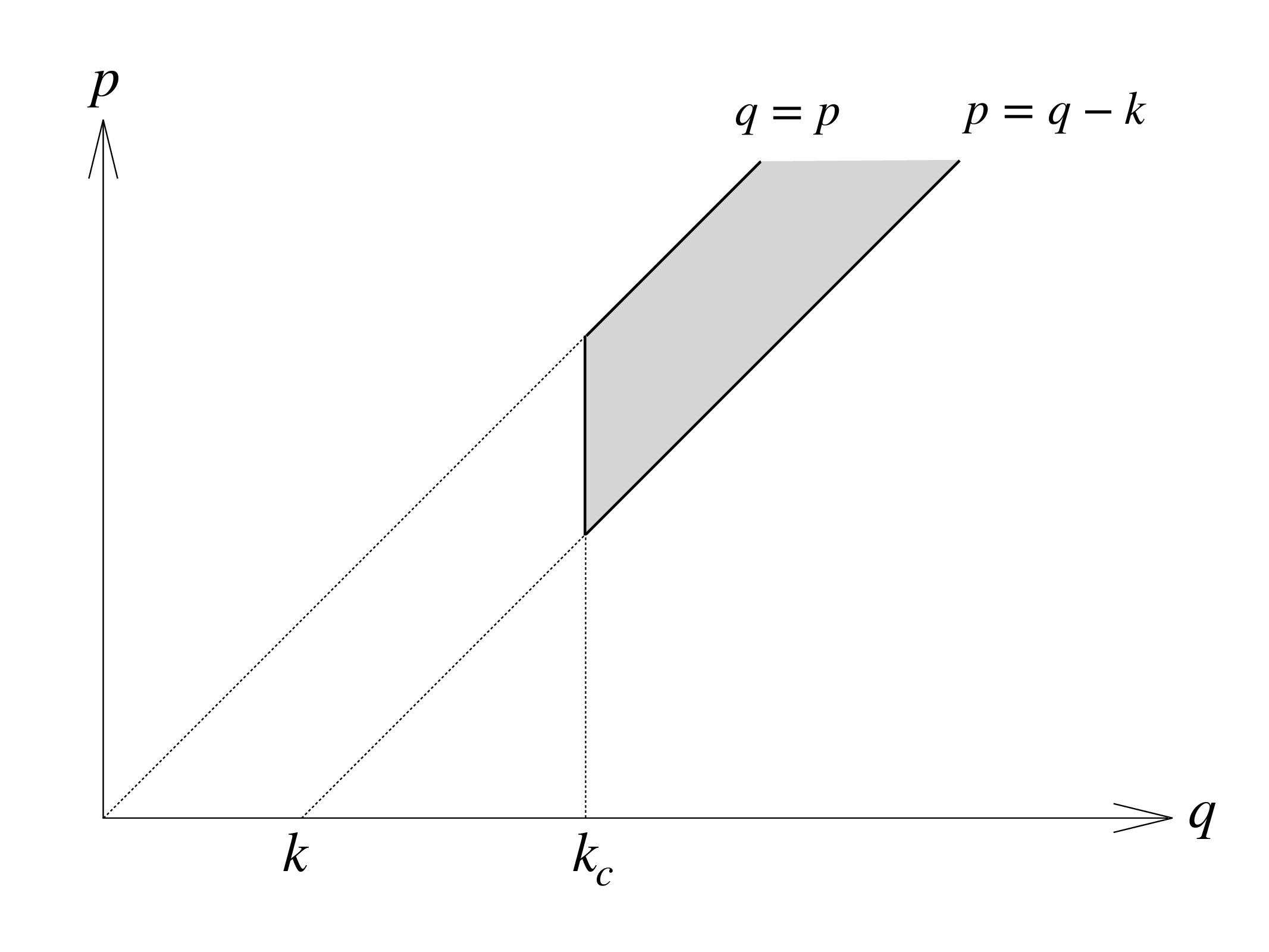}
\caption{For symmetry of $S_\theta(k;p,q)=S_\theta(k;q,p)$ in the scalar-transport function $T_\theta(k,t)$, the above gray area may be sufficient in calculating the high-wavenumber contribution of Eq. (\ref{Ttheta0}). }
\label{high_T_area}
\end{figure}

Non-local interaction can be further understood using so called eddy-diffusivity concept. Let us focus on the high-wavenumber contribution to the transport function $T_\theta(k,t)$ of Eq. (\ref{Ttheta0}) (see Fig. \ref{high_T_area}); setting a cutoff wavenumber $k_c$, one may calculate the contribution from $\triangle \cup\{(p,q)|p,q>k_c\gg k \}$, which may be written as $T_\theta^{>k_c}(k,t)$.  In case of $H(k)\gg H(p),H(q)$ we reach its limit
\begin{equation}
T_\theta^{>k_c}(k,t) = -2\kappa_T(k_c)k^2 E_\theta(k,t)\  \  \  (k/q \to 0),
\end{equation}
where
\begin{equation}
\kappa_T(k_c) =\frac{4}{3}\pi \int^\infty_{k_c} dq \int^t_{-\infty} ds\ q^2 Q(q;t,s)G_H(q;t,s).
\end{equation}
Note that $\kappa_T(k_c)$ is independent from $k$, exactly providing the eddy-diffusivity concept where the scalar variance diffuses proportionally to its second-order derivative. Then HBLRA tells us that the eddy diffusivity originates from fluid's random motion and memory fading of the geometry of the scalar distribution (Hessian) in small scale $q\gg k$, which is due to geometrical basis of our formulation. Conventional result by Ref. \citep{Kaneda86} is consistently obtained under a replacement $G_H(q;t,s)\to G_\theta(q;t,s)$, where the memory fading of the scalar is absent. Suppose that $k$, $k_c$, and $q$ are all in the inertial-convective range, we have %, which is identical to that of Ref. \citep{Kraichnan68} except for the inertial-range solution of $Q(q;t,s)$.
\begin{equation}
\kappa_T(k_c) \approx 0.746 \bepsilon^{1/3} k_c^{-4/3}, 
\end{equation}
which is smaller than that of scalar-based LRA; i.e.  $1.02 \bepsilon^{1/3} k_c^{-4/3}$ Ref. \citep{Kaneda86}. Then the weaker non-local coupling in HBLRA may be further interpreted as the lower eddy-diffusivity. Introducing the eddy viscosity $\nu_T(k_c)$ in a similar way \citep{Kraichnan65,Kaneda86}, we can define the Schmidt number from these effective viscosity and diffusivity:
\begin{equation}
\frac{\nu_T(k_c)}{\kappa_T(k_c)} \approx 0.30. 
\end{equation}
which is now larger than $0.22$ of scalar-based LRA .\\

While we have so far compared the scalar-based LRA with HBLRA, similar comparisons may be made by other closure theories; i.e.  ALHDIA, SBALHDIA, and RI-LRA, all of which do not incorporate the timescale of scalar itself, represent stronger interference from large-scale eddy just like scalar-based LRA does. Not only the quantitative difference in the universal Obukhov-Corrsin constant, but there are essential discrepancies between present and conventional theories in their physical background, which ultimately arrive back to the incorporation of timescale of scalar itself.

%One should note that this does not contradicts Kolmogorov-Obukhov-Corrsin analysis since the interscale interaction is still localized within the inertial range, but may be more broadband than the 

\section{Conclusion}
In the present work, we have constructed a closure theory HBLRA (Hessian-based LRA) for the passive scalar turbulence on the basis of the Hessian field of the scalar. Unlike the Lagrangian scalar, the Hessian statistics firmly represents the memory-fading effect caused by random turbulence motion, which essentially changes the quantitative prediction on scalar statistics. Following the systematic LRA procedures, a self-consistent closure model is derived for second-order statistics of the Hessian field. The resultant closure model has realistic features as a physical model such as the detailed conservation (\ref{d balance}) and scale-locality of the non-linear interacion, all of which are the very key to consistency with the Kolmogorov-Obukhov-Corrsin theory. In particular, the scale-locality of the nonlinear self-interaction strictly regulates possible candidates for representatives, and our Hessian field $\mathscr{H}_{ij}$ successfully meets this physical requirement.\\

It is worth mentioning the differences between the current HBLRA and pioneering RI-LRA as another branch of LRA \cite{GNK00}. Let us make below comparisons of the two theories from (i) physical and (ii) practical aspects. 

\emph{(i) physical aspect} -- RI-LRA employs pure strain tensor besides the scalar field so that the memory fading of the scalar is accounted by the time scale of the pure straining motion, while the current HBLRA introduces the time scale of the Hessian of the scalar itself; to be brief, RI-LRA focuses on the velocity field, while HBLRA on the scalar field. Also RI-LRA, sharing its core idea with SBALHDIA, is specifically designed for physics in high-wavenumber region, focusing on the strain field associated with the spectral transfer in small scale range \cite{GNK00,KH78}, so we cannot tell now about its performance in larger-scale ranges, e.g., inertial-convective and inertial-diffusive ranges.

\emph{(ii) practical aspect} -- In HBLRA, the representative variables to be solved are the Lagrangian correlation function $H$ and response function $G_H$ of the Hessian of the scalar field and the correaltion function $Q$ of the velocity field. Regarding the inertial-convective-range analysis, $Q$ had been already obtained in Ref. \cite{Kaneda86}, so $H$ and $G_H$ can be solved without difficulties just like in Sec. \ref{INERTIAL CONVECTIVE}. On the other hand, RI-LRA treats the Lagrangian correlation and response functions of the strain field and those ($\Theta$ and $G_\theta$) of the scalar field as its representative variables. Although the latter two may be trivially solved for their lack in the memory-fading effect (see Sec. \ref{SCALAR FIELD}), the closure of the former strain statistics has not been achieved yet for mathematical complexities of RI-LRA formalism. Instead, the strain statistics are estimated with the help of short-time analysis and DNS data in Ref. \cite{GNK00}. Then fundamental assessments (solvability of the closure equations, consistency with demensional analyses, etc.) are yet to be done for RI-LRA before its practical applications.\\

The inertial-convective range has been investigated on the basis of HBLRA, where the known Kolmogorov-Obukhov-Corrsin scaling has been obtained as the solution of the theory. Our self-consistent approach now derives the Obukhov-Corrsin constant $K_\theta$ reasonably close to DNS and experimental values ranging from 0.6 to 0.9 \cite{Sreeni96, MW98, WCB99, YXS02, WG04, GW15}. Whereas we have not yet reached conclusive arguments to decide what is the best choice for representative, our three candidates, i.e. primary, solenoidal, and traceless Hessians, may have simplest mathematical significance among any possibilities (see Sec. \ref{HESSIAN FIELD}), which helps us to access their physical interpretations in easy-to-understand ways. In particular, the solenoidal Hessian may be the most suitable choice in properly representing local scalar structure (see discussions in Sec. \ref{HESSIAN FIELD}), and its resultant $K_\theta\approx 0.754$ well agrees with what is obtained by Ref. \cite{GW15} ($K_\theta\approx 0.725$). On the other hand, recent large-scale DNS \cite{Ishihara16} suggests that the \emph{true} inertial range, if exists, could be observed at the scale even larger than $100\eta$. Thence the true validation of Obukhov-Corrsin theory, which also relies on the classical Kolmogorov scaling, may be postponed until even larger scale DNS in future studies.\\

%while the recent large-scale DNS by Ref. \cite{} suggests $K_\theta\approx 0.7$ which is close to the result of solenoidal Hessian $K_\theta = 0.725$.

Besides the inertial-convective scaling, there are still many aspects to be explored by HBLRA; physics in the inertial-diffusive and viscous-convective ranges where other scaling laws are expected as universal features of high and low P\'{e}clet-number problems subjected to high-Reynolds-number turbulence \cite{Goto-Kida99}. In addition to stationary cases, time-dependent numerical simulations of HBLRA may further explicate dynamical features of scalar turbulence; e.g. passive scalar stirred by decaying turbulence or decay of the scalar itself. Not only for simply passive cases, but HBLRA may extend its applicability to more general scalar field subjected to active interactions; buoyancy-driven turbulence, turbulent chemical reaction, and turbulent particle-clustering \cite{AYMY18} may be typical examples where theoretical supports may be even more needed to account for their complex physics. We believe these applications in future studies may lead us to more sophisticated understanding of scalar turbulence with the help of true governing law.\\\\ 

\noindent\textbf{Acknowledgments}

We were benefited from valuable discussions with Profs. Yukio Kaneda and Katsunori Yoshimatsu, which greatly helps us to brush up the current work. Also we would like to acknowledge fruitful communication with Prof. Ye Zhou and his encouraging comments on the present work. This study was supported by JSPS Grant-in-Aid for Scientific Research (S) JP16H06339.

%HBLRA, as a selfconsistent physical model, ca

%decaying laws and large-scale invariance, etc. 

%Not only passive, but also active scalars will be the target of current study. Besides isotropic cases, scalar subjected to fully anisotropic turbulence will be also an important issue.

\def\theequation{\Alph{section}\textperiodcentered\arabic{equation}}
\appendix

\section{Lagrangian position function}\label{POSITION FUNCTION}
Although the Lagrangian picture offers an essential platform to some fundamental understandings of fluid mechanics, it often brings about complexities in the actual treatment of physical properties. Regarding this general issue, Lagrangian position function enables us to bridge the gap between the Lagrangian and Eulerian pictures by a feasible manner \cite{Kaneda81}. The Lagrangian position function $\psi(\mathbf{x}',t;\mathbf{x},t')$ is defined as a density function of the Lagrangian tracer particle located on $\mathbf{x}$ at time $t'$. Then, for an arbitrary physical property $\phi(\mathbf{x},t)$ in the Eulerian representation, one may introduce its Lagrangian counterpart by 
\begin{equation}
\phi(\mathbf{x},t'|t)=\int d^3x'\psi(\mathbf{x}',t;\mathbf{x},t')
\phi(\mathbf{x}',t),
\end{equation}
where the Lagrangian field $\phi(\mathbf{x},t'|t)$ is the value of $\phi$ experienced by the Lagrangian tracer particle passing through the space-time point $(\mathbf{x},t')$. For a given turbulent velocity field $\mathbf{u}(\mathbf{x},t)$, $\psi(\mathbf{x}',t;\mathbf{x},t')$ is governed by 
\begin{equation}
\partial_t \psi(\mathbf{x}',t;\mathbf{x},t')
+\partial'_j\left[ u_j(\mathbf{x}',t)\psi(\mathbf{x}',t;\mathbf{x},t')\right]=0
\end{equation}
accompanied by its initial condition $\psi(\mathbf{x}',t';\mathbf{x},t')$ $=$ $\delta^3(\mathbf{x}'-\mathbf{x})$. Now $\psi$ can be treated as a field in Eulerian picture, so one can access the Lagrangian quantities in a similar manner to the Eulerian analysis.

\section{Eulerian and Lagrangian velocity fields}\label{EULER-LAGRANGE VELOCITY}
In the present study, the Eulerian and Lagrangian velocity field in Fourier space are governed by the following set of equations accompanied by Eq. (\ref{psi 1}):
\begin{equation}
\left(\partial_t+\nu k^2\right)u^{\scalebox{0.6}{E}}_i(\mathbf{k},t)
=\lambda\frac{1}{i}M_{i.ab}(\mathbf{k})[\mathbf{k;p,q}]
u^{\scalebox{0.6}{E}}_a(\mathbf{p},t)u^{\scalebox{0.6}{E}}_b(\mathbf{q},t),
\label{EU eq.}
\end{equation}
\begin{equation}
\left(\partial_t+\nu k^2\right)G^{\scalebox{0.6}{E}}_{ij}(\mathbf{k},t;\mathbf{k}',t')
=2\lambda\frac{1}{i}M_{i.ab}(\mathbf{k})[\mathbf{k;p,q}]
G^{\scalebox{0.6}{E}}_{aj}(\mathbf{p},t;\mathbf{k}',t')u^{\scalebox{0.6}{E}}_b(\mathbf{q},t) \ (t\geq t'),
\label{EGU eq.}
\end{equation}
\begin{equation}
\begin{split}
\partial_t u^{\scalebox{0.6}{L}}_i(\mathbf{k},t'|t)
=&\lambda\int d^3k'' \psi(\mathbf{k}'',t;\mathbf{k},t')
\frac{ik''_i k''_a k''_b}{k''{}^2}
[\mathbf{k}'';\mathbf{p,q}]u^{\scalebox{0.6}{E}}_a(\mathbf{p},t)u^{\scalebox{0.6}{E}}_b(\mathbf{q},t)\\
&-\nu\int d^3k'' \psi(\mathbf{k}'',t;\mathbf{k},t') k''{}^2 
u^{\scalebox{0.6}{E}}_i(\mathbf{k}'',t),
\end{split}
\label{LU eq.}
\end{equation}
\begin{equation}
\begin{split}
\partial_t G^{\scalebox{0.6}{L}}_{ij}(\mathbf{k},t;\mathbf{k}',t')
=&2\lambda\int d^3k'' \psi(\mathbf{k}'',t;\mathbf{k},t')
\frac{ik''_i k''_a k''_b}{k''{}^2}
[\mathbf{k}'';\mathbf{p,q}]u^{\scalebox{0.6}{E}}_a(\mathbf{p},t)G^{\scalebox{0.6}{E}}_{bj}(\mathbf{q},t;\mathbf{k}',t')\\
&+\lambda\int d^3k'' \Psi_j(\mathbf{k}'',t;\mathbf{k};\mathbf{k}',t')
\frac{ik''_i k''_a k''_b}{k''{}^2}
[\mathbf{k}'';\mathbf{p,q}]u^{\scalebox{0.6}{E}}_a(\mathbf{p},t)u^{\scalebox{0.6}{E}}_b(\mathbf{q},t)\\
&-\nu\int d^3k'' \psi(\mathbf{k}'',t;\mathbf{k},t') k''{}^2 
G^{\scalebox{0.6}{E}}_{ij}(\mathbf{k},t;\mathbf{k}',t')\\
&-\nu\int d^3k'' \Psi_j(\mathbf{k}'',t;\mathbf{k};\mathbf{k}',t')k''{}^2 
u^{\scalebox{0.6}{E}}_i(\mathbf{k},t;\mathbf{k}',t') \ (t\geq t'),
\end{split}
\label{LGU eq.}
\end{equation}
%\begin{equation}
%\partial_t\psi(\mathbf{k}'',t;\mathbf{k},t')
%=\lambda ik''_b[\mathbf{k}'';\mathbf{-p,q}]u_b(\mathbf{p},t)\psi(\mathbf{q},t;\mathbf{k},t'),
%\label{psi 2}
%\end{equation}
\begin{equation}
\begin{split}
\partial_t\Psi_j(\mathbf{k}'',t;\mathbf{k};\mathbf{k}',t')
=&ik''_a[\mathbf{k}'';\mathbf{-p,q}]G^{\scalebox{0.6}{E}}_{aj}(\mathbf{p},t;\mathbf{k}',t')\psi(\mathbf{q},t;\mathbf{k},t'),\\
&+ik''_a[\mathbf{k}'';\mathbf{-p,q}]u^{\scalebox{0.6}{E}}_a(\mathbf{p},t)\Psi_j(\mathbf{q},t;\mathbf{k};\mathbf{k}',t')\ (t\geq t'),
\end{split}
\label{psi-response}
\end{equation}
\begin{equation}
G^{\scalebox{0.6}{E}}_{ij}(\mathbf{k},t';\mathbf{k}',t')
=G^{\scalebox{0.6}{L}}_{ij}(\mathbf{k},t';\mathbf{k}',t')
=P_{ij}(\mathbf{k})\delta^3(\mathbf{k}-\mathbf{k}'),
\label{velocity-response initial}
\end{equation}
\begin{equation}
\Psi_j(\mathbf{k}'',t';\mathbf{k};\mathbf{k}',t')=0,
\label{psi-response initial}
\end{equation}
where $M_{i.ab}$ ($\equiv (P_{ia}k_b+P_{ib}k_a)/2$) is another solenoidal operator, $\Psi_j$ is the response of the position function against disturbance applied to the velocity field. Non-perturbative counterparts of the above read
\begin{equation}
\left(\partial_t+\nu k^2\right)\tilde{u}_i(\mathbf{k},t)=0,
\end{equation}
\begin{equation}
\left(\partial_t+\nu k^2\right)\tilde{G}_{ij}(\mathbf{k},t;\mathbf{k}',t')=0\ (t\geq t'),
\end{equation}
\begin{equation}
\tilde{G}_{ij}(\mathbf{k},t';\mathbf{k}',t')=
P_{ij}(\mathbf{k})\delta^3(\mathbf{k}-\mathbf{k}'),
\end{equation}
\begin{equation}
\tilde{\Psi}_j(\mathbf{k}'',t;\mathbf{k};\mathbf{k}',t')=0.
\label{np Psi}
\end{equation}
Applying the renormalization procedure of Sec. \ref{LAGRANGIAN RENORMALIZATION} on the basis of Eqs. (\ref{EU eq.})-(\ref{np Psi}), we reach the closure model for $Q_{ij}(\mathbf{k};t,t')$ and $G_{ij}(\mathbf{k};t,t')$ identical to Eqs. (2.35)-(2.46) of Ref. \cite{Kaneda81}.

\section{Renormalization procedure}\label{RENORMALIZATION}
To summarize the core idea of the renormalization, we employ below some symbolic notations; we denote correlation tensors only by their main symbol; e.g. $\mathcal{H}$ stands for $\mathcal{H}_{ij.lm}(\mathbf{k};t,t')$. According to Ref. \cite{Kaneda81}, an arbitrary correlation can be closed in the following steps:\\

\noindent (i) Assume both velocity and scalar fields at initial time instant $t_0$, i.e. $u_i(\mathbf{k},t_0)$ and $\theta(\mathbf{k},t_0)$, to be all independent Gaussian random, where the non-perturbatibve solutions ($\lambda$-zeroth-order solutions) of velocity and Hessian, say $\tilde{u}_i(\mathbf{k},t)$ and $\tilde{\mathscr{H}}_{ij}(\mathbf{k},t)$, become also independent Gaussian random. Then an arbitrary correlation, say $J$, can be expanded in terms of the correlation of the non-perturbative variables; i.e. $\tilde{\mathcal{H}}$, $\tilde{\mathcal{G}}$, $\tilde{Q}$, and $\tilde{G}$:
\begin{equation}
J=\mathfrak{J}^{\scalebox{0.7}{(0)}}[\tilde{\mathcal{H}},\tilde{\mathcal{G}},\tilde{Q},\tilde{G}]
+\lambda \mathfrak{J}^{\scalebox{0.7}{(1)}}[\tilde{\mathcal{H}},\tilde{\mathcal{G}},\tilde{Q},\tilde{G}]+O(\lambda^2),
\label{expanded J}
\end{equation}
where $\mathcal{J}^{\scalebox{0.7}{(n)}}[\cdots]$ are functionals. \\

\noindent (ii) In the same manner, expand $\mathcal{H}$, $\mathcal{G}$, $Q$, and $G$ in terms of $\tilde{\mathcal{H}}$, $\tilde{\mathcal{G}}$, $\tilde{Q}$, and $\tilde{G}$:
\begin{equation}
\begin{split}
\mathcal{H}&=\tilde{\mathcal{H}}
+\lambda^2 \mathfrak{A}^{\scalebox{0.7}{(2)}}[\tilde{\mathcal{H}},\tilde{\mathcal{G}},\tilde{Q}]+O(\lambda^4),\\
\mathcal{G}&=\tilde{\mathcal{G}}
+\lambda^2 \mathfrak{B}^{\scalebox{0.7}{(2)}}[\tilde{\mathcal{H}},\tilde{\mathcal{G}},\tilde{Q}]+O(\lambda^4),\\
Q&=\tilde{Q}
+\lambda^2 \mathfrak{C}^{\scalebox{0.7}{(2)}}[\tilde{Q},\tilde{G}]+O(\lambda^4),\\
G&=\tilde{G}
+\lambda^2 \mathfrak{D}^{\scalebox{0.7}{(2)}}[\tilde{Q},\tilde{G}]+O(\lambda^4),
\end{split}
\label{primitive expansion}
\end{equation}
where $\mathfrak{A}^{\scalebox{0.7}{(n)}}[\cdots]$, $\mathfrak{B}^{\scalebox{0.7}{(n)}}[\cdots]$, $\mathfrak{C}^{\scalebox{0.7}{(n)}}[\cdots]$, and $\mathfrak{D}^{\scalebox{0.7}{(n)}}[\cdots]$ are all functionals (odd orders vanish for the Gaussianity).\\

\noindent(iii) Invert Eqs. (\ref{primitive expansion}) (To be precise, there is an ambiguity in the reverse-expansion technique, which may be mentioned later.):
\begin{equation}
\begin{split}
\tilde{\mathcal{H}}&=\mathcal{H}
-\lambda^2 \mathfrak{A}^{\scalebox{0.7}{(2)}}[\mathcal{H},\mathcal{G},Q]
+O(\lambda^4),\\
\tilde{\mathcal{G}}&=\mathcal{G}
-\lambda^2 \mathfrak{B}^{\scalebox{0.7}{(2)}}[\mathcal{H},\mathcal{G},Q]
+O(\lambda^4),\\
\tilde{Q}&=Q
-\lambda^2 \mathfrak{C}^{\scalebox{0.7}{(2)}}[Q,G]+O(\lambda^4),\\
\tilde{G}&=G
-\lambda^2 \mathfrak{D}^{\scalebox{0.7}{(2)}}[Q,G]+O(\lambda^4),
\end{split}
\label{reverse expansion}
\end{equation}\\

\noindent(iv) By substituting Eqs. (\ref{reverse expansion}) into Eq. (\ref{expanded J}), $J$ can be expressed by representative variables $\mathcal{H}$, $\mathcal{G}$, $Q$, and $G$:
\begin{equation}
J=\mathfrak{J}^{\scalebox{0.7}{(0)}}[\mathcal{H},\mathcal{G},Q,G]
+\lambda \mathfrak{J}^{\scalebox{0.7}{(1)}}[\mathcal{H},\mathcal{G},Q,G]+O(\lambda^2).
\label{renormalized J}
\end{equation}\\

\noindent(v) Truncate the renormalized expansion (\ref{renormalized J}) at the lowest order.\\

For instance, let the exact dynamical equation of $\mathcal{H}$ be $\partial_t\mathcal{H}=\lambda I$ ($I$ is a triple correlation). Now the renormalization procedure (i)-(v) yields 
\begin{equation}
I=\lambda \mathfrak{I}^{\scalebox{0.7}{(1)}}[\mathcal{H},\mathcal{G},Q]
+O(\lambda^2)
\overset{\mathrm{truncation}}{\longrightarrow} \lambda \mathfrak{I}^{\scalebox{0.7}{(1)}}[\mathcal{H},\mathcal{G},Q],
\label{renormalized I}
\end{equation}
which then results in
\begin{equation}
\partial_t\mathcal{H}=\lambda^2 \mathfrak{I}^{\scalebox{0.7}{(1)}}[\mathcal{H},\mathcal{G},Q].
\end{equation}
Likewise, dynamical equations for all the representatives can be closed by themselves. 
So far we have followed the renormalization procedure of Ref. \cite{Kraichnan77} originally introduced by Kraichnan. Here we shall focus on the problem hidden behind the difference between Eulerian and Lagrangian renormalization. Unlike in the Eulerian formalism, Lagrangian correlation $\mathcal{H}_{ij.lm}(\mathbf{k};t,t')$ is asymmetric in its Eulerian and Lagrangian labels:
\begin{equation}
\mathcal{H}_{ij.lm}(\mathbf{k};t,t')
\neq \mathcal{H}_{lm.ij}(-\mathbf{k};t',t),
\end{equation}
which may cause an uncertainty in the renormalization procedure. Simple perturbation expansions of them read
\begin{subequations}
\begin{align}
\mathcal{H}(t,t')=\tilde{\mathcal{H}}(t,t')+\lambda^2\mathfrak{A}[\tilde{\mathcal{H}},\tilde{\mathcal{G}},\tilde{Q}](t,t')+O(\lambda^4),\\
\mathcal{H}(t',t)=\tilde{\mathcal{H}}(t',t)+\lambda^2\mathfrak{A}[\tilde{\mathcal{H}},\tilde{\mathcal{G}},\tilde{Q}](t',t)+O(\lambda^4) .
\end{align}
\end{subequations}
Then their reverse expansions are given by
\begin{subequations}
\begin{align}
\tilde{\mathcal{H}}(t,t')=\mathcal{H}(t,t')-\lambda^2\mathfrak{A}[\mathcal{H},\mathcal{G},Q](t,t')+O(\lambda^4),\label{forward expansion}\\
\tilde{\mathcal{H}}(t',t)=\mathcal{H}(t',t)-\lambda^2\mathfrak{A}[\mathcal{H},\mathcal{G},Q](t',t)+O(\lambda^4).\label{backward expansion}
\end{align}
\end{subequations}
Note that left sides of Eqs. (\ref{forward expansion}) and (\ref{backward expansion}) take the same value ($\tilde{\mathcal{H}}(t,t')=\tilde{\mathcal{H}}(t',t)$ by definition), while the finite truncation of right sides results in their difference. In case of the lowest-order truncation, this is equivalent to replace $\tilde{\mathcal{H}}(t,t')$ with either of $\mathcal{H}(t,t')$ or $\mathcal{H}(t',t)$, where no evident selection rule exists so far. Fortunately, at the lowest-order renormalization, every $\mathcal{H}$ directly inherits one time variable from either of Lagrangian or Eulerian equations, so we can avoid the ambiguity given above. This is how we conducted the Lagrangian renormalization in Sec. \ref{LAGRANGIAN RENORMALIZATION}.\\

%For example, regarding the second term of the right side of Eq. (\ref{2TC eq.0}), Lagrangian time $t$ of $H(-\mathbf{k};t,t')$ originates from the expansion of the Lagrangian Hessian, while the remaining $t'$ does from the Eulerian Hessian. 

%Note that renormalized quantities are to be selected so that sub group of infinite series expansions are properly replaced by some finite terms. Also note that all the bare terms in primitive perturbation expansion appear from Eulerian quantities, which realizes a time-symmetric expansion of the reduceable terms absorbed into renormalized two-time correlation. This fact implies that the renormalized two-time correlation should be symmetric under the exchange of $t$ and $t'$.

In more general case, however, the ambiguity problem cannot be removed by such a simple algorithm. Here we see another renormalization procedure of wider applicablity for future progress of Lagrangian closures. The ambiguity exactly comes from the time-asymmetry of the Lagrangian auto-correlation. Thus we choose its time-symmetric part as an alternative representative to it; i.e.
\begin{equation}
\mathcal{H}^{\scalebox{0.6}{S}}(t,t')
\equiv\frac{1}{2}\left[ 
\mathcal{H}(t,t')+\mathcal{H}(t',t)
\right],
\end{equation}
where the superscript $S$ stands for time-symmetrization. Likewise we did for $\mathcal{H}$ in Sec. \ref{LAGRANGIAN RENORMALIZATION}, $\mathcal{H}^{\scalebox{0.6}{S}}$ can be expanded as
\begin{equation}
\mathcal{H}^{\scalebox{0.6}{S}}(t,t')
=\tilde{\mathcal{H}}(t,t')+\lambda^2\mathfrak{A}^{\scalebox{0.6}{S}}[\tilde{\mathcal{H}},\tilde{\mathcal{G}},\tilde{Q}](t,t')+O(\lambda^4),
\end{equation}
which is now \emph{uniquely} reverted as
\begin{equation}
\tilde{\mathcal{H}}(t,t')
=\mathcal{H}^{\scalebox{0.6}{S}}(t,t')-\lambda^2\mathfrak{A}^{\scalebox{0.6}{S}}[\mathcal{H}^{\scalebox{0.6}{S}},\mathcal{G},Q](t,t')+O(\lambda^4).
\end{equation}
Now the reverse expansion uniquely determine the renormalization at an arbitrary order of truncation. Then Eqs. (\ref{ETC LRA'}) and (\ref{2TC LRA'}) become
\begin{equation}
\begin{split}
&\left(\partial_t+2\kappa k^2\right)H(\mathbf{k};t,t)\\
&=2[\mathbf{k};\mathbf{p},\mathbf{q}]\int^t_{t_0}ds\, 
Q(\mathbf{p};t,s)
\left\{k_ak_b H^{\scalebox{0.6}{S}}(\mathbf{q};s,t)G_H(-\mathbf{k};t,s)
-k_aq_b H^{\scalebox{0.6}{S}}(-\mathbf{k};s,t)G_H(\mathbf{q};t,s)\right\},
\end{split}
\label{ETC sLRA}
\end{equation}

\begin{equation}
\begin{split}
\left(\partial_t+\kappa k^2\right)&H(\mathbf{k};t,t')\\
=&-[\mathbf{k};\mathbf{p},\mathbf{q}]
X_{ab}(\mathbf{k},\mathbf{q})Z_{ab}(\mathbf{k})q_cq_d \int^t_{t'} ds Q_{cd}(\mathbf{p};t,s) H^{\scalebox{0.6}{S}}(-\mathbf{k},t,t')\\
&-[\mathbf{k};\mathbf{p},\mathbf{q}]X_{ab}(\mathbf{k},\mathbf{q})Z_{ab}(\mathbf{k})k_cq_d\int^t_{t_0}ds\, 
Q_{cd}(\mathbf{p};t,s)G_H(\mathbf{q};t,s) H^{\scalebox{0.6}{S}}(-\mathbf{k};s,t')\\
&+[\mathbf{k};\mathbf{p},\mathbf{q}]X_{ab}(\mathbf{k},\mathbf{q})Z_{ab}(\mathbf{k})k_ck_d\int^{t'}_{t_0}ds\, 
Q_{cd}(\mathbf{p};t,s) H^{\scalebox{0.6}{S}}(\mathbf{q};t,s)G_H(-\mathbf{k};t',s),
\end{split}
\label{2TC sLRA}
\end{equation}
while Eq. (\ref{G LRA'}) is not to be altered. Following the same procedure in Sec. \ref{INERTIAL CONVECTIVE}, we reach the Obkhov-Corrsin constant:
\begin{equation}
K_\theta\approx
\begin{dcases}
&0.989\ (\textrm{primary Hessian}),\\
&0.754\ (\textrm{solenoidal Hessian}),\\
&1.18\ (\textrm{traceless Hessian}),\\
\end{dcases}
\end{equation}
which slightly differs from the previous result of Eq. (\ref{OC constant}).\\

Exactly the same problem can be discussed in the velocity closure. In general, Lagrangian velocity correlation may be asymmetric under the exchange of the Eulerian and Lagrangian labels:
\begin{equation}
Q_{ij}(\mathbf{k};t,t')\neq Q_{ji}(-\mathbf{k};t',t).
\end{equation}
In Ref. \cite{Kaneda81}, the uncertainty problem is avoided by restricting the domain of $Q_{ij}(\mathbf{k};t,t')$ to $t\geq t'$ (also see Eq. (\ref{renormalized quantities})), which is technically adoptable due to a fortunate situation of the velocity closure; LRA equation for $Q_{ij}(\mathbf{k};t,t')$ requires only $[t',t]$ as its integration domain. In more general cases (e.g. ALHDIA, SBALHDIA, and HBLRA), however, one needs wider domain of time integration where we cannot restrict the domain of definition (ALHDIA and SBALHDIA define their Lagrangian correlation for arbitrary two times by a discontinuous exchange between the Eulerian and Lagrangian labels). To develop more unified understanding of the theories, one can apply the above symmetric renormalization to the velocity closure, which brings about some modifications to the original LRA equations outside the inertial range.

%The symmetric renormalization can be applied to any Lagrangian closure theories described above, which may remove an ambiguity problem  

%(see the second terms on the right sides of Eqs. (\ref{2TC eq.0}), (\ref{2TC LRA'}), (\ref{HI 2TC eq.}), and (\ref{2TC sLRA}))

%In such integration, the time-ordering restriction may be

\section{Scale-similar representation of the spectral flux}\label{SCALE-SIMILAR PI}
Due to the detailed conservation (\ref{d balance}), the integration (\ref{Pi0}) over the range $\{(k',p',q')|k',p',q'>k\}$ vanishes. Then Eq. (\ref{Pi0}) is reduced to
\begin{equation}
\Pi_\theta(k,t)=\int^\infty_k dk' 
\int^k_0 dp'\int^{p'+k'}_{\mathrm{max}(p',k'-p')} dq'\, S_\theta(k';p',q'|t).
\label{Pi}
\end{equation}
In the inertial-convective range, a scale-similar transformation $S_\theta(k';p',q'|t)=a^3 S_\theta(ak';ap',aq'|t)$ holds. Here we choose $a=k/k'$, $b=ap'$, and $q=aq'$, so that Eq. (\ref{Pi}) becomes
\begin{equation}
\Pi_\theta(k,t)=\int^\infty_k dk' \frac{k}{k'}
\int^{k^2/k'}_0 db\int^{b+k}_{\mathrm{max}(b,k-b)} dq\, S_\theta(k;b,q|t),
\label{Pi1}
\end{equation}
By setting $p=k^2/k'$ Eq. (\ref{Pi1}) turns into
\begin{equation}
\Pi_\theta(k,t)=\int^k_0 dp\ \frac{k}{p}
\int^p_0 db\int^{b+k}_{\mathrm{max}(b,k-b)} dq\, S_\theta(k;b,q|t).
\label{Pi2}
\end{equation}
Partial integration by $p$ yields
\begin{equation}
\begin{split}
\Pi_\theta(k,t)=&\Bigg[k\,\mathrm{ln}\left(\frac{p}{k}\right)
\int^p_0 db\int^{b+k}_{\mathrm{max}(b,k-b)} dq\, S_\theta(k;b,q|t)\Bigg]^k_0
-\int^k_0 dp\, k\,\mathrm{ln}\left(\frac{p}{k}\right)
\int^{p+k}_{\mathrm{max}(p,k-p)} dq\, S_\theta(k;p,q|t)\\
=&k\int^k_0 dp\, \mathrm{ln}\left(\frac{k}{p}\right)
\int^{p+k}_{\mathrm{max}(p,k-p)} dq\, S_\theta(k;p,q|t).
\end{split}
\end{equation}
which coincides with Eq. (\ref{Pi3}).

\end{document}